\documentclass[14pt]{article}
\pdfoutput=1
\usepackage{amsmath,amssymb,amsfonts,amsthm,bm,bbm,cancel,wasysym}
\usepackage{epsfig,graphics,graphicx,epstopdf}
\usepackage{array,booktabs,colortbl,colordvi,multirow}
\usepackage{colordvi,color,xcolor}
\usepackage{hyperref}
\usepackage{rotating}

\usepackage{verbatim}
\usepackage{cite}
\usepackage{subfig}
\usepackage{setspace}
\usepackage{url}
\usepackage[percent]{overpic}
\usepackage{slashed}
\usepackage{authblk}
\usepackage{xspace}
\usepackage{fullpage}
\usepackage{hyperref}

\def\ie{{\it i.e.}}
\def\eg{{\it e.g.}}
\def\etc{{\it etc}}

\def\to{\rightarrow}

\newskip\zatskip \zatskip=0pt plus0pt minus0pt
\def\matth{\mathsurround=0pt}
\def\lsim{\mathrel{\mathpalette\atversim<}}
\def\gsim{\mathrel{\mathpalette\atversim>}}
\def\atversim#1#2{\lower0.7ex\vbox{\baselineskip\zatskip\lineskip\zatskip
  \lineskiplimit 0pt\ialign{$\matth#1\hfil##\hfil$\crcr#2\crcr\sim\crcr}}}

\begin{document}

%----------------------------------- TITLE AND AUTHORS -----------------------------------------%

%Preprint numbers
\begin{flushright}
SLAC-PUB-17576\\
\today
\end{flushright}
\vspace*{5mm}

\renewcommand{\thefootnote}{\fnsymbol{footnote}}
\setcounter{footnote}{1}

\begin{center}

{\Large {\bf The Bactrian Effect: Multiple Resonances and Light Dirac Dark Matter}}\\
%\vspace*{0.15cm}

\vspace*{0.75cm}

{\bf Thomas G. Rizzo}~\footnote{rizzo@slac.stanford.edu}

\vspace{0.5cm}

{SLAC National Accelerator Laboratory}\ 
{2575 Sand Hill Rd., Menlo Park, CA, 94025 USA}

\end{center}
\vspace{.5cm}

%--------------------------------------------- ABSTRACT ---------------------------------------------%

\begin{abstract}
\noindent                                                                                                                                
The possibility of light dark matter (DM) annihilating through a dark photon (DP) which kinetically mixes (KM) with the Standard Model (SM) hypercharge field is a very attractive 
scenario. For DM in the interesting mass range below $\sim 1$ GeV, it is well known that bounds from the CMB provide a very strong model building constraint forcing the DM annihilation 
cross section to be roughly 3 orders of magnitude below that needed to reproduce the observed relic density.  Under most circumstances this removes the possibility of an $s$-wave 
annihilation process for DM in this mass range as would be the case, \eg,  if the DM were a Dirac fermion. In an extra-dimensional setup explored previously, it was found that the 
$s$-channel exchange of multiple gauge bosons could simultaneously encompass a suppressed annihilation cross section during the CMB era while also producing a sufficiently large 
annihilation rate during freeze-out to recover the DM relic density. In this paper, we analyze more globally the necessary requirements for this mechanism to work successfully 
and then realize them within the context of a simple model with two `dark'  gauge bosons having masses of a similar magnitude and whose contributions to the annihilation amplitude 
destructively interfere. We show that if the DM mass threshold lies appropriately in the saddle region of this destructive interference between the two resonance humps it then becomes 
possible to satisfy these requirements simultaneously provided several ancillary conditions are met. The multiple constraints on the parameter space of this setup are then explored in detail 
to identify the phenomenologically successful regions.                                                                                                              
\end{abstract}

\renewcommand{\thefootnote}{\arabic{footnote}}
\setcounter{footnote}{0}
\thispagestyle{empty}
\vfill
\newpage
\setcounter{page}{1}

%-------------------------------- DOCUMENT: INTRODUCTION ---------------------------------%

% 1 Introduction

\section{Introduction}

Although dark matter (DM) is known to exist at multiple scales in the universe we don't yet know what it is or if it interacts with the particles of the Standard Model (SM) through any forces 
other than via gravity. However, in order to obtain the observed relic density as measured by Planck\cite{Aghanim:2018eyx} it is more than likely that some sort of non-gravitational 
interactions are responsible. The traditional DM candidates, Weakly Interacting Massive Particles (WIMPs)\cite{Arcadi:2017kky,Roszkowski:2017nbc} and the familiar  
axion\cite{Kawasaki:2013ae,Graham:2015ouw,Irastorza:2018dyq}, either assume the usual Standard Model (SM) electroweak interactions or some new high scale physics is 
responsible for obtaining the relic density. While such theories remain very interesting, the lack of any observational signatures at the LHC or in either direct or indirect detection 
searches\cite{LHC,Aprile:2018dbl,Fermi-LAT:2016uux,Amole:2019fdf} has resulted in a slowly shrinking allowed parameter space for these models. This has led to the construction of a 
plethora of new DM scenarios based on the introduction of non-SM interactions to reproduce the observed relic abundance\cite{Steigman:2015hda,Saikawa:2020swg} with very wide 
ranges in both the possible DM masses and coupling strengths\cite{Alexander:2016aln,Battaglieri:2017aum,Bertone:2018krk}. Many of these potential new interactions can 
be described via a set of `portals' which link DM, and possibly other `dark' sector fields, with those of the SM, only a few of which can result from renormalizable, dimension-4 terms in 
the Lagrangian. 

Perhaps the most attractive of these ideas, and one that has received much attention in the recent literature,  is the vector boson/kinetic mixing (KM) portal\cite{KM,vectorportal} which 
will be the subject of the analysis that follows below. The main ingredients of this setup in its basic incarnation can be deceptively simple: DM is assumed to be a SM singlet but 
instead carries a charge under a new `dark' gauge interaction, \eg, $U(1)_D$, with a corresponding gauge coupling $g_D$. The associated gauge field is thus termed the `dark photon' (DP) 
\cite{Fabbrichesi:2020wbt} which has a mass that can be generated by the dark analog of the usual Higgs mechanism, \ie, via the `dark Higgs'. The coupling of the DM and other dark 
sector fields to the SM is then generated by the KM of the $U(1)_D$ DP with the SM $U(1)_Y$ hypercharge gauge boson which can be accomplished at 1-loop via a set of 
`portal matter' fields that are charged under both gauge groups\cite{Rizzo:2018vlb,Rueter:2019wdf,Kim:2019oyh,Wojcik:2020wgm,Rueter:2020qhf}. Once all the fields are canonically 
normalized to remove the effects of this KM and both the $U(1)_D$ and SM gauge symmetries are spontaneously broken, one finds that the the DP has picked up a small 
loop-induced coupling to the SM fields. For the range of DP masses below $\sim 1$ GeV that we will consider in our analysis, to leading order in the DP to SM $Z$ mass-squared ratio, 
one finds the well-known result that this coupling can be very well approximated as $\simeq \epsilon eQ_{em}$, where 
$\epsilon$ is a dimensionless parameter, here assumed to roughly lie in the 
interval $\sim 10^{-4} -10^{-3}$, that describes the magnitude of this loop-suppressed KM. 

When both the DM and the DP are both light and have somewhat comparable masses, $\lsim 1$ GeV, the DM can still be a thermal relic in a manner similar to what happens in the 
conventional WIMP scenario. The proximity of these two masses can occur naturally in several setups: for example, if a common dark Higgs vev generates both the DM mass 
and is simultaneously responsible for the breaking of $U(1)_D$ or in KM models with extra dimensions where the compactification radius sets the common scale for particle masses 
\cite{Rizzo:2020ybl,Landim:2019epv,Rizzo:2018joy,Rizzo:2018ntg}. In this low mass regime, there are several constraints on the model parameters: first, there is the required 
annihilation cross section necessary to obtain the observed relic density during freeze out, \eg, $<\sigma \beta_{rel}>_{FO} \simeq 4.5\times 10^{-26}$ cm$^3$s$^{-1}$ (here $\beta_{rel}$ 
is the relative DM velocity in the collision process) for an 
$s$-wave annihilating Dirac fermion DM\cite{Steigman:2015hda,Saikawa:2020swg}, which is the case that we will consider below. For such a light mass, we will assume in what 
follows that pair annihilation of DM via virtual spin-1 exchanges is responsible for this and that it results in a SM final state consisting of pairs of electrons, muons, or light charged hadrons. 
Second, a lower bound on the DM mass exists arising from Big Bang Nucleosynthesis considerations of roughly $\sim 10$ MeV (which we take from Ref.\cite{Sabti:2019mhn}). 
Lastly, in this same DM mass range of $\sim 10-1000$ MeV, the CMB (at $z\sim 10^3$) constraints from Planck\cite{Aghanim:2018eyx} tell us that at that time the DM annihilation cross 
section into light SM charged states, \eg, $e^+e^-$, must be substantially suppressed\cite{Slatyer:2015jla,Liu:2016cnk,Leane:2018kjk,Bringmann:2006mu} thus avoiding the 
possible injection of any 
additional electromagnetic energy into the SM plasma. A recent analysis\cite{Cang:2020exa} of this constraint informs us that it lies roughly at the level of  
$\sim 5 \times 10^{-29}~(m_{DM}/100~ {\rm {MeV}})$ cm$^3$s$^{-1}$, noting that it depends approximately linearly on the DM mass, but is, in any case, roughly three orders of magnitude 
below that needed at freeze out to recover the observed relic density. However, as the DM get heavier, this constraint becomes quite weak and can be essentially ignorable for DM 
masses above roughly $\gsim 10-20$ GeV.  We further note that this constraint from the CMB is not expected to strengthen by more than a factor of $\sim 2$ in the coming 
years\cite{Green:2018pmd,Ade:2018sbj,Abazajian:2016yjj}.  There are also constraints of a very similar magnitude for this range of DM masses from a completely different source which 
are found to arise from Voyager 2 data\cite{Boudaud:2016mos,Boudaud:2018oya}.

These simultaneous requirements pose a strong set of constraints on the nature of DM and how it may annihilate into the SM via the $s$-channel exchange of spin-1 mediators like the 
DP, \eg, if DM is a Dirac fermion (as will be considered here), this annihilation process is dominantly $s$-wave assuming vector couplings. In such a case, since the reaction rate is 
generally not very sensitive to the relative velocity of the annihilating DM, $\beta_{rel}$,  the cross sections at freeze-out and during the CMB are not expected to be much different thus 
conflicting with the requirements above. Does this imply that light Dirac fermion DM in the KM setup and annihilating to the SM as described above is excluded in this mass range? In 
the simple canonical DP scenario -- without any `tweaking' -- as discussed earlier, the answer in `yes'. However, modifications of this basic idea may allow for this possibility and several 
more or 
less successful but diverging paths might be followed, one of which we will consider here. In recent work\cite{Rizzo:2018joy} on the 5-D extension of this usual 4-D KM setup with Dirac DM,  
it was found in a random scan that certain regions of the model parameter space simultaneously satisfied the CMB bound while still leading to the desired DM annihilation cross section 
(via multiple $s$-channel Kaluza-Klein DP exchanges) at freeze out. While the exact mechanism at work in this case was speculated upon and the necessary ingredients for this success 
never fully identified, it was clear that the existence of more than one particle exchange and with the proper interference structure were clearly necessary ingredients. In this paper, we will 
further examine this issue in some detail and then construct a simpler, more tractable and transparent 4-D scenario which satisfies all of the necessary conditions. To this end we will 
employ a modified version of the dark sector model considered in Ref.\cite{Rueter:2019wdf} based on a SM-like, but fully broken, $SU(2)_I\times U(1)_{Y_I}$ dark gauge group, 
naturally having two diagonally coupled gauge bosons with only a few adjustable parameters. 

The outline of this paper is as follows: in Section 2, based on our previous work, we consider and outline in detail the necessary inputs and constraints on a model of Dirac fermion DM which 
interacts with the fields of the SM through (at least) a pair of two spin-1 mediators, $Z_i$, whose couplings are generated by KM, thus generalizing the conventional DP setup. We then 
construct a simple but realistic model that satisfies all of these requirements. In Section 3, we discuss the phenomenological implications of the model we construct based on the 
requirements arrived at in the previous Section and then we explore how their interplay impacts the model's parameter surviving space. Our results and conclusions are then summarized 
in Section 4.

%-------------------------------- DOCUMENT: SECTIONS -----------------------------------------%

\section{Model Setup and Basics}

In this Section, we will discuss the essential requirements for and the set of constraints imposed upon models that may realize the expectations described above as well as the 
reasoning behind them. A simple, prototypical -- but potentially physically realistic -- proof of principle model of this kind with the desired properties will then be presented and 
examined in some detail. 

\subsection{Model Building Constraints} 

The mechanism envisioned here has several important distinct components -- some of which were superficially touched upon in our earlier work\cite{Rizzo:2018joy}. In this subsection 
we will clarify what these are and what their interplay is with one another. Based on these observations we will make a number of model building assumptions in what follows and then 
explore how they can be realized. 

($i$) We imagine that light Dirac fermionic DM, $\chi$, with a mass in the 10 to 1000 MeV range, realizes the observed relic density via the usual pair annihilation to SM fields, \eg, 
$e^+e^-$, via the $s$-channel exchange of two (or more) new neutral gauge bosons, $Z_i$, which have masses of comparable magnitude to the DM. As noted above, this is a 
rather natural occurrence in, \eg, ED models of KM wherein the masses of all the low lying states are set by the inverse size of the extra size of the ED, 
$R^{-1}$\cite{Rizzo:2020ybl,Landim:2019epv,Rizzo:2018joy,Rizzo:2018ntg}, or in models where the masses are determined by the single 
vacuum expectation value (vev) of a scalar field. While the couplings of the DM to the $Z_i$ will be set by a common overall dark gauge coupling, $g_D$ (modulo Clebsch-Gordon and 
mixing angle factors as appear in the SM), the $Z_i$ couplings to SM fields will be determined up to similar overall factors via a single kinetic mixing with the SM hypercharge field and so 
are related to one another but are also, as is usual,  suppressed by loop factors. This loose framework is just a rather straightforward generalization of the familiar DP/KM 
model\cite{KM,vectorportal}. To simplify matters and make things more tractable we will specifically concern ourselves with the case of only two $Z_i\to Z_{1,2}$ in what follows but 
the arguments we make can be generalized as in the case of, \eg,  ED that was previously considered as well as to other scenarios with multiple $s$-channel spin-1 exchanges.

($ii$) Though axial-vector couplings of the DM to the $Z_i$, $a_i^{DM}$,  can be present, and we will return to this possibility below, we will assume that the DM [and the SM fermions] 
must at least have [only have] vector couplings to the $Z_i$, $v_i^{DM,SM}\neq 0$,  so that in the non-relativistic, low relative velocity limit, $\beta_{rel}^2 \to 0$,  the annihilation process 
is primarily an 
$s$-wave and is also not, \eg, helicity or threshold suppressed by any small SM fermion masses that may appear in the final state. This is, again, just a generalization of the familiar 
DP/KM scenario. 

\begin{figure}[htbp]
\centerline{\includegraphics[width=5.0in,angle=0]{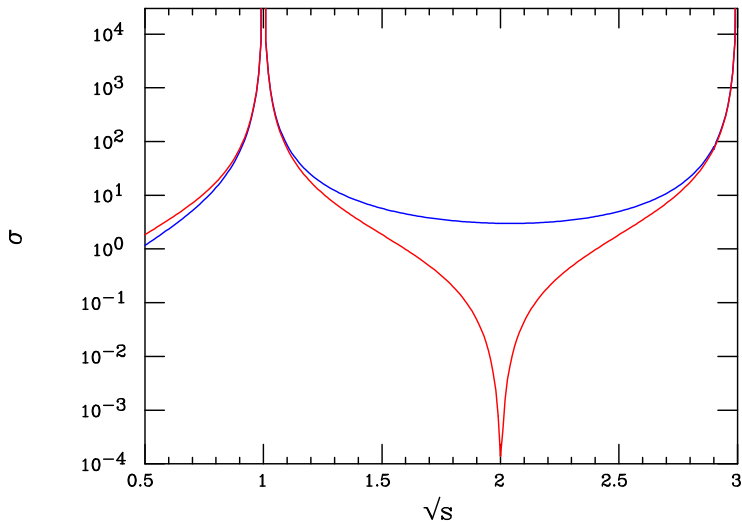}}
\vspace*{-1.50cm}
\caption{Semi-quantitative picture of the Dirac fermion DM annihilation cross section, in random units, when $\sqrt s$ is given in units of $m_1$ and where $m_2/m_1=3$ 
has been assumed for purposes of demonstration. Both the constructive or destructive interference possibilities are shown.} 
\label{sample}
\end{figure}

($iii$) Now consider the DM annihilation cross section during the CMB era at $z\sim 10^3$ when the temperature is sufficiently low so that taking the $\beta_{rel}^2 \to 0$ approximation is an 
excellent one and thus any axial-vector couplings of the DM to the $Z_i$,  $a_i^{DM}$,  can be safely ignored as their contributions to the annihilation rate are all $\beta_{rel}^2$ suppressed 
in this case. This implies that in this limit there is effectively only a single `vector-vector' coupling amplitude contributing to the DM annihilation process to a given final fermion state 
which is made up of the sum of the individual contributions of the various $Z_i$ and which we can write in the simple familiar form
\begin{equation}
A \simeq \sum_{i=1,2} \frac{v_i^{DM}v_i^{SM}}{s-m_i^2+i\Gamma_im_i}\,,  
\end{equation}
where $s$ is the usual Mandelstam variable and here $m_i=m_{Z_i}$ with $\Gamma_i$ being the total widths of these states (assumed here to be at least somewhat narrow 
$\Gamma_i/m_i <$ a few $\%$ or perhaps significantly smaller). Further, we now make the additional assumption that the DM mass is such that $2m_{DM}$ lies within the saddle region, \ie,    
$m_1 < 2m_{DM}=\sqrt s < m_2$ between the two resonance humps where the equality follows from the fact that we have taken $\beta_{rel}^2 \to 0$;  such a situation may be envisioned as 
that shown qualitatively in Fig.~\ref{sample}. Trivially, if the product of the DM and SM couplings to the $Z_{1,2}$ have the same (opposite) sign in both cases, then destructive (constructive) 
interference between the two contributions to the amplitude in the saddle region between the resonances will take place. In the case of destructive interference, in which we will be 
interested, the location of this very deep cross section minimum lies approximately (in the zero width limit) at the center of mass energy 
\begin{equation}
\sqrt s_0 \simeq \Big[\frac{m_2^2+X^2m_1^2}{1+X^2}\Big]^{1/2}\,,  
\end{equation}
where $X^2>0$ is the ratio of the product of the couplings of two the $Z_i$, \ie, $X^2 = v_2^{DM}v_2^{SM}/(v_1^{DM}v_1^{SM})$. For a fixed value of the mass ratio $m_2/m_1$, the value 
of $X^2$ determines the proximity of this minimum to the location of either resonance, \eg, moving closer to $m_1$ relative to $m_2$ as $X^2$ increases. (We will return to this 
relationship below within the context of a specific model.) Now we easily imagine that for DM lying in this mass range this destructive interference is at least partially responsible for the 
relatively suppressed annihilation cross section which must hold during the CMB (as well at at present times), provided the value of $m_{DM}$ is properly chosen. 

While such a 
deep destructive interference may be possible to achieve if two or more distinct amplitude structures of comparable magnitude contributed to the annihilation process, it certainly would 
be significantly more difficult to arrange since the precise relative weights of the contributions to the total amplitude would in general be quite different{\footnote {For example, while there 
can be destructive interference of the $\gamma$ and $Z$ contributions below the SM $Z$ resonance for the familiar $e^+e^- \to \bar ff$ process, the resulting cross section suppression 
is not extremely large due to the existence of several competing amplitudes.}}.  This is our reasoning behind the assumption made above that the $Z_i$ couplings to the SM are solely 
vector-like; while taking the $\beta_{rel}^2 \to 0$ limit allows us to `turn off' any contribution to the total amplitude from DM axial-vector couplings during the CMB, it generally cannot `turn off' 
those associated with SM axial-vector couplings. 

It goes without saying that the necessary suppression of the annihilation cross section at the time of the CMB in such a setup 
will result in a similarly suppressed annihilation rate today by factors of $\sim 10^3$ (as it is an $s$-wave process)  in comparison to usual expectations. This implies that we should 
expect no observable DM annihilation signals in present indirect detection experiments with rates anywhere close to the conventionally assumed $\sim 10^{-26}$  cm$^3$s$^{-1}$ value.

So far we have been considering `SM' in the above to be a single entity, \eg, in the mass range of interest to us here it may simply be the $e^+e^-$ final state. Of course if the DM is 
sufficiently massive then other final states such as $\mu^+\mu^-$ and/or hadrons may also be kinematically accessible and the total annihilation cross section is then a weighted sum of these 
various contributions. It is, of course, a strongly destructive minimum in this {\it total} cross section that we seek here. In such a case, certainly, we will need all of these individual contributions 
to have destructive minima at the same value of $\sqrt s$ as given by the expression above and to that end we must require that the ratio $v_2^{SM}/v_1^{SM}$ be the same for all accessible 
SM final states. Note that this is a {\it weaker} requirement than demanding that the separate $v_{1,2}^{SM}$ {\it individually} be the same for all of these final states. This weaker requirement 
can be easily satisfied if, \eg, $v_i^{SM}=c_iQ_{em}$ (or with $Q_{em}$ here replaced by any other fixed combination of gauge group generators), where the $c_i$ are final state 
independent constants. This will indeed be the case in the simple model that we will construct below and this requirement occurs relatively naturally if both of 
these couplings are generated via the same KM but result in different corresponding strengths simply due to mixing angle effects.

In the case with two $Z_i$, the lightest one lies in the mass regime where in it will decay (with an $\epsilon^2$ suppressed rate) exclusively to visible SM states while the somewhat 
heavier $Z_2$  is allowed to have unsuppressed decays to DM and so be `invisible', \eg, in accelerator experiments. This would imply that such a scenario would necessarily 
simultaneously lead to {\it both} types of signals that are usually discussed when looking for dark photon production. Since both types of searches would need to be satisfied, depending 
upon the relative masses and couplings of these two states, one could then (appropriately) simultaneously combine the constraints from both classes of searches in a correlated manner 
to constrain the parameter space of such a model. Such an analysis lies beyond the scope of the current discussion but in the simple model below we will choose parameters so that 
these searches are {\it individually} satisfied ignoring any correlations.

($iv$) A further constraint on this setup is that we must require (in its weakest form and again something we will return more seriously to below) that $m_{DM}<m_1$ so that the $s$-wave, 
non-KM {\it or} velocity suppressed process $\bar \chi \chi \to 2Z_1$ is kinematically forbidden when $\beta_{rel}^2\to 0$ otherwise 
the DM annihilation rate during the CMB will very easily violate the bound above. As will be discussed below, finite temperature effects, \eg, during freeze-out, will lead to a required 
strengthening of this bound as can be gleaned from the detailed studies of Forbidden DM models\cite{Griest:1990kh,DAgnolo:2015ujb,Cline:2017tka,Fitzpatrick:2020vba,1837855}.
As we will see below, this also leads to a further bound on the ratio $m_2/m_1$ and thus will also play a rather strict role as a constraint on our model parameter space.

($v$) Although we may manage to sufficiently suppress the annihilation rate of Dirac fermion DM during the CMB via destructive interference, we still need to have a correspondingly 
large annihilation cross section at freeze-out, $<\sigma \beta_{rel}>_{FO}\simeq 4.5 \cdot 10^{-26}$ cm$^3$s$^{-1}\equiv 4.5\sigma_0$\cite{Steigman:2015hda,Saikawa:2020swg},  for Dirac 
fermion DM when $m_{DM}/T_{FO} \simeq 20$ or so, to recover the DM relic density as 
observed by Planck. To do this we rely on the non-zero temperature effects present during the early universe to insure that $\beta_{rel}^2 \neq 0$ be large enough to sufficiently raise 
the center of mass energy for at least some of the DM collisions. Thus, as in the case of resonant enhancement, we imagine that with $T_{FO}\neq 0$ the DM has a 
sufficiently enhanced center of mass energy to feel the influence of the $Z_2$ resonance hump. For the mass range of interest to us here, this effect must be strong enough  
so as to enhance the annihilation cross section in comparison to CMB times by a factor by roughly $K\sim $ a few $\cdot 10^3$ or so as mentioned above and will be further discussed  
below. However, unlike in the case of ordinary resonant enhancement, the cross section in our case starts out quite suppressed at low temperatures due to 
the destructive interference implying that these finite temperature effects may now be potentially much more significant as we saw in our earlier work on ED. Obviously, if $m_2$ is too  
large in comparison to $2m_{DM}$ 
the influence of this second resonance will be reduced unless the coupling ratio $X^2$ is sufficiently large so as to compensate for this effect. We note that due to ($iv$) we cannot 
arbitrarily increase the value of $m_{DM}$ to bring the DM `closer' to experiencing the second hump and, since need to rely only on these thermal effects, $m_2$ cannot be made arbitrarily 
large in comparison to $m_1$. Thus we might expect that, \eg, $m_2/m_1 \lsim 3$ or even less is necessary to make this approach effective. Clearly this balance of potentially 
conflicting constraints will require some detailed numerical study within a specific framework to determine if they can be simultaneously satisfied. To address all of these issues we 
now consider a rather simple, but physically interesting, proof of principle toy model wherein each can be examined in turn.

\subsection{Constructing a Simple Model} 

To move forward, we consider a simple model of the dark sector gauge interactions a variant of which we have analyzed previously\cite{Rueter:2019wdf} in a very different context and 
which we will realize here in a somewhat different manner. Consider generalizing the familiar the dark gauge group from $U(1)_D$ to $SU(2)_I \times U(1)_{Y_I}$ with the gauge 
couplings $g_I,g_I'$ in analogy with the SM. Unlike in the earlier version of this model, here the SM fields themselves will remain singlets under this gauge group. 
Unlike in the SM, however, this gauge group must be {\it completely} broken at or below the $\lsim 1$ GeV mass scale; as will be seen below 
this complete symmetry breaking requires the action of (at least) two `dark' Higgs multiplets acquiring vevs to supply the required Goldstone bosons. In analogy with the SM, we can define 
a corresponding set of quantities $g_I'/g_I=\tan \theta_I=t_I$, $e_I=g_Is_I=g_I'c_I$, with $s_I=\sin \theta_I$, \etc. Note that, again analogous to the SM, we will define the `dark charge' 
to which the dark photon would couple as $Q_D=T_{3I}+Y_I/2$ in familiar SM-like notation. 

It is convenient to begin this discussion by first considering the KM between the SM $U(1)_Y$ 
hypercharge gauge boson, $\hat B_\mu$, and the analogous $U(1)_{Y_I}$ field, $\hat B_I^\mu$, generated as usual at the 1-loop level through the action of some portal matter (PM) 
fields but whose detailed nature is beyond the scope of the present discussion\cite{Rizzo:2018vlb,Rueter:2019wdf,Kim:2019oyh,Wojcik:2020wgm,Rueter:2020qhf}. 
This KM is described in familiar notation by
\begin{equation}
{\cal L}_{KM} =\frac{\epsilon}{2c_wc_I} \hat B_{\mu\nu} \hat B_I^{\mu\nu}\,,  
\end{equation}
where typically $\epsilon=10^{-(3-4)}$. Here we will always consider $\epsilon$ to be sufficiently small so that we can generally work to linear order in this parameter except where 
necessary. This KM is removed (to lowest order in $\epsilon$) via the usual simple field redefinitions: $\hat B\to B+\frac{\epsilon}{c_wc_I} B_I$ and $\hat B_I\to B_I$.
Now consider all of the gauge fields in the SM plus those in the dark sector in a familiar basis:  $W^\pm$, $Z$ and $A$ defined as usual and now also 
$W_I^\pm$ (where here the $\pm$ labels the electrically neutral $W_I$'s {\it dark} charge as we will see below), $\hat Z_I$ and $\hat A_I$.  In such a basis, after KM has been removed, 
the SM gauge fields will couple as they usually do but the hermitian dark sector gauge fields will pick up additional interactions proportional to the SM hypercharge
\begin{equation}
\frac{g_I}{\sqrt 2} T_I^+W_I +h.c. +e_IQ_DA_I +\frac{g_I}{c_I} (T_{3I}-s_I^2Q_D) Z_I+\frac{\epsilon g_Y}{c_wc_I} \frac{Y}{2} (c_IA_I-s_IZ_I)\,, 
\end{equation}
Note that at this point we have only removed the KM and have gone to a somewhat convenient and familiar basis; {\it none} of the gauge symmetries have yet been broken which is what 
we need to do next.

As usual, we will assume that the SM gauge group is broken by the $T_{3L}=-Y/2=1/2$ vev, $v\simeq 246.2$ GeV, of a weak isodoublet which carries no dark quantum numbers and 
gives, \eg, the $W^\pm$ it's usual tree-level mass, $M_W=(gv/2)$, while leaving the SM photon massless. Of course $v\neq 0$ also generates the usual diagonal mass term for the $Z$, 
$M_Z^2=(gv/2c_w)^2$, but, via the KM terms in the couplings, there will also be both diagonal and off-diagonal terms in the dark sector as well as mixing terms with the $Z$. We note, 
however, that at this step dark gauge symmetries remain unbroken. To accomplish this further breaking we first add an $SU(2)_I$ doublet, SM singlet scalar field which has $Y_I/2=-1/2$ 
and whose $Q_D=0$ element obtains a vev, $v_D \sim1$ GeV; this generates a mass for $W_I^\pm$, \ie, $M_{W_I}=g_Iv_D/2$, in analogy with the SM. Second, we add an additional 
$SU(2)_I$, as well as SM, singlet complex scalar field with $Q_D=1$ that also obtains a vev, $v_S$, of a similar (but perhaps slightly smaller) magnitude. Abbreviating 
the suggestive combinations $M_{Z_I}^2=(g_Iv_D/2c_I)^2$ and $M_{A_I}^2=(e_Iv_S)^2$, the full $3\times 3$ neutral gauge boson mass squared matrix (the SM photon remaining 
massless and decouples, of course) then becomes in the $(Z,A_I,Z_I)$ basis
\begin{equation}
M_{3 \times 3}^2 =
\begin{pmatrix}
M_Z^2 & -\epsilon t_w M_Z^2 & \epsilon t_wt_I M_Z^2\\ 
-\epsilon t_w M_Z^2 & \epsilon^2t_w^2t_I^2M_Z^2+M_{A_I}^2 & -\epsilon^2 t_wt_I M_Z^2-t_I M_{A_I}^2 \\ 
\epsilon t_wt_I M_Z^2 & -\epsilon^2 t_wt_I M_Z^2-t_I M_{A_I}^2 & \epsilon^2t_w^2t_I^2M_Z^2+t_I^2M_{A_I}^2+M_{Z_I}^2
\end{pmatrix}.
\end{equation}
Making the small rotations $A_I\to A_I-\epsilon t_wZ$, $Z_I\to Z_I+\epsilon t_wt_I Z$ and $Z\to Z+\epsilon t _w(A_I-t_IZ_I)$ then removes the mixings between the now physical $Z$ 
and both $A_I,Z_I$ to this order as well as all of $O(\epsilon^2)$ entries in the lower right $2\times 2$ submatrix. Combining these results with Eq.(4) above, some algebra tells us that 
the $A_I,Z_I$ gauge bosons (which are not yet mass eigenstates) will now couple to SM fields in the combination $e\epsilon Q_{em}(A_I-t_I Z_I)$ and that the physical $Z$ picks up 
an $O(\epsilon)$ 
coupling to the dark sector fields. These results assume that $M_{Z_I,A_I}^2<<M_Z^2$ which is certainly true for the parameter choices we have made so far. We can now decouple the 
$Z$ and then the remaining neutral gauge boson mixing is seen to lie totally within the dark sector and has significantly simplified to just (now in the $A_I,Z_I$ basis): 
\begin{equation}
M_{2 \times 2}^2 =
\begin{pmatrix}
M_{A_I}^2 & -t_I M_{A_I}^2\\ 
-t_I M_{A_I}^2 & t_I^2M_{A_I}^2+M_{Z_I}^2
\end{pmatrix},
\end{equation}
where we now see very transparently that the $Q_D=1$ singlet vev, $v_S$, is obviously required for both of the eigenstates masses to be non-zero. This matrix is easily diagonalized by 
defining the new mass eigenstate fields $Z_{1,2}$ where $A_I=Z_1 c_\phi- Z_2 s_\phi$ and $Z_I=Z_2 c_\phi+Z_1 s_\phi$ with $s_\phi (c_\phi)=\sin \phi (\cos \phi)$ and where the angle 
$\phi$ is given by the expression 
\begin{equation}
\tan 2\phi =\frac{2t_I M_{A_I}^2}{M_{Z_I}^2+(t_I^2-1)M_{A_I}^2}\,.
\end{equation}
In terms of the physical fields $Z_{1,2}$, the coupling of these dark gauge bosons with the visible sector SM can be simply written as
\begin{equation}
{\cal L}_{SM-int}=e\epsilon_{eff}Q_{em}(Z_1-TZ_2)\,,
\end{equation}
where we have now defined the combinations 
\begin{equation}
T=\tan(\phi+\theta_I)=\frac{t_\phi+t_I}{1-t_\phi t_I}, ~~~~~~\epsilon_{eff}=\epsilon(c_\phi-t_Is_\phi)=\epsilon c_\phi(1-t_\phi t_I)\,.
\end{equation}
Note that, within the parameter ranges to be employed below, it is always true that $\epsilon_{eff} \leq \epsilon$. 
Also note that, trivially, the $Z_{1,2}$ couplings to the SM are proportional to one another, \ie, 
$v_1^{SM}=e\epsilon_{eff}Q_{em}$, $v_2^{SM}=-Tv_1^{SM}$ in the notation of Eq.(1).  The corresponding couplings of the $Z_i$ to the dark sector fields are given by 
\begin{equation}
{\cal L}_{DM-int}=\Big[\frac{g_I}{c_I}(T_{3I}-s_I^2Q_D)s_\phi+e_IQ_Dc_\phi\Big]Z_1+\Big[\frac{g_I}{c_I}(T_{3I}-s_I^2Q_D)c_\phi-e_IQ_Ds_\phi\Big]Z_2\,.
\end{equation}
To go further we must posit the transformation of the DM field under $SU(2)_I \times U(1)_{Y_I}$ requiring, trivially, that $Q_D(\chi)\neq 0$ and that the DM be the lightest member of 
the $SU(2)_I$ multiplet to which it belongs to insure its stability. The simplest possibility satisfying these requirements is that $\chi$ is a $Q_D=1$ state which is also an $SU(2)_I$ 
singlet, \ie, $T_{3I}(\chi)=0$~{\footnote{The dark sector may, of course, contain other additional fields in various multiplets of the dark gauge symmetry all of which are more massive 
than the DM itself.}}. 
Assuming this to be the case, then if we define the combination 
\begin{equation}
g_D=e_Ic_\phi(1-t_\phi t_I)\,,
\end{equation}
we obtain that $v_1^\chi=g_DQ_D(\chi)\equiv g_D$ and $v_2^\chi=-Tv_1^\chi$. Finally, combining both sets of couplings we observe that
\begin{equation}
\frac{v_2^\chi v_2^{SM}}{v_1^\chi v_1^{SM}}= T^2\,,
\end{equation}
where we see that we've reproduced the desired result from the discussion in the previous subsection above with the identification 
$X^2\to T^2$ and, since $T=\tan(\phi+\theta_I)$, $0\leq T^2\leq \infty$. 

Next, we need to address the masses of the $Z_i$ themselves, $m_i$, and their relationships to the other model parameters. Given the discussion in the previous subsection we recall 
that we will be particularly interested in parameter values where the mass ratio $\lambda_R= m_2/m_1$ is held fixed. Given the simple form of the mass squared matrix above it is clear that 
the ratio of its eigenvalues, $\lambda_R^2=\lambda_+/\lambda_-$, will depend only upon the value of $t_I$ and the ratio $\rho=M_{A_I}^2/M_{Z_I}^2$. Explicitly,
\begin{equation}
(M^2_{Z_I})^{-1} \lambda_\pm =\frac{1}{2} \big[1+\rho (1+t_I^2)\big] \pm \frac{1}{2} \big[1+2\rho (t_I^2-1)+\rho^2(1+t_i^2)^2\big]^{1/2}\equiv A\pm B\,,
\end{equation}
so that 
\begin{equation}
\lambda_R^2=\frac{1+R}{1-R} ~~~{\rm with} ~~~ R=\frac{B}{A}=\frac{\lambda_R^2-1}{\lambda_R^2+1} \,.
\end{equation}
For a given $\lambda_R$ one can now determine (the physical) value of $\rho(t_I)$ as the `+' root of quadratic equation 
\begin{equation}
R^2-1+2\big[R^2(1+t_I^2)+1-t_I^2\big]\rho+(R^2-1)(1+t_I^2)^2\rho^2=0\,,
\end{equation}
and requiring this root to be real places an upper bound on $t_I$:
\begin{equation}
t_I^{max}=\frac{R}{(1-R^2)^{1/2}}=\frac{\lambda_R^2-1}{2\lambda_R}\,,
\end{equation}
with $\rho(t_I^{max})=[1+t_I^{max~2}]^{-1}$. Using the definition of the angle $\phi$ in terms of $\rho$ and $t_I$ then leads to an analogous 
upper bound on $t_\phi$ which after some algebra becomes 
\begin{equation}
t_\phi^{max}=\big[1+t_I^{max~2}\big]^{1/2}-t_I^{max}=\lambda_R^{-1}\,,
\end{equation}
so that, after more algebra and employing the definition of $T$ above, we finally arrive at the simple upper bound
\begin{equation}
T^{max}=\lambda_R\,.
\end{equation}
This bound is phenomenologically very important because, as we noted above, we will 
need to increase $T$ as $m_2/m_1$ becomes larger to keep the cross section minimum within the range given by the requirements ($iv$) and ($v$) above. 

To see how this parameter constraint and the other requirements above play out in this setup, we need to perform a detailed numerical study to which we now turn.

\section{Bactrian Phenomenology} 

This model as constructed has only vectorial couplings for the DM and SM to the $Z_i$ and basically has only 3 dimensionless parameters apart from an overall coupling strength and 
a mass scale; we take these parameters to be $r=2m_{DM}/m_1$, $\lambda_R=m_2/m_1$ and $T$.  As we saw above, model consistency plus 
phenomenological constraints will likely impose somewhat sever restrictions on their interrelated allowed values. 

To proceed, we first consider the DM annihilation cross section for the process $\bar \chi \chi \to Z_i^* \to \bar f f$, where the fermion field, $f$, is here being used as 
a placeholder for the SM in generality. This cross section is given in the above model by by a simple generalization of the well-known result\cite{Berlin:2014tja}
\begin{equation}
\sigma \beta_{rel} = \frac{2\alpha \epsilon_{eff}^2g_D^2}{3s} N_c^f Q_f^2 ~\beta_f \frac{3-\beta_f^2}{2} ~\frac{3-\beta_\chi^2}{2} ~\sum_{i,j} P_{ij} \tilde v_i^\chi \tilde v_j^\chi \tilde v_i^f \tilde v_j^f\,,
\end{equation}
where $\beta_{\chi,f}^2=1-4m_{\chi,f}^2/s$. {\footnote {The presence of possible additional axial couplings of the DM to the $Z_i$ can be easily accommodated by letting (in a common 
normalization)  $\tilde v_i^\chi \tilde v_j^\chi\to \tilde v_i^\chi \tilde v_j^\chi+2\beta_\chi^2(\tilde a_i^\chi \tilde a_j^\chi)/(3-\beta_\chi^2)$ in this expression above. However,  
this does not happen in the present simple model realization that we are considering here but if present would 
generally only make O(1) modifications to the discussion below at the time of freeze-out but would have no effect during the CMB as noted previously.}} 
As is also clear, and as previously noted, the $\beta_\chi^2$ terms will essentially vanish at the time of the CMB due to the low temperatures/DM velocities. 
For simplicity, we will consider the specific case of $f=e$ in what follows so that $N_c^f=|Q_f|=\beta_f=1$ in the kinematic region of interest but the reader should remember that the 
cross section may be a factor of a few time larger numerically when additional final state channels become kinematically allowed. 
Note that with this chosen normalization and employing the results above we find that $\tilde v_1^\chi=\tilde v_1^f=1$ and $\tilde v_2^\chi=\tilde v_2^f=-T$. The kinematic propagator 
factor appearing in this expression,  $P_{ij}$, is given as usual by
\begin{equation}
P_{ij}= s^2 ~\frac{(s-m_i^2)(s-m_j^2)+ \Gamma_i\Gamma_j m_im_j}{[(s-m_i^2)^2+(\Gamma_im_i)^2][i ~\to ~j]}\,.
\end{equation}
Since we are assuming that $m_1<2m_{DM}<m_2$ as per the above discussion, $Z_1$ can decay only to SM states, \ie, the electron, so that it has a suppressed width, 
$\Gamma_1/m_1=(e\epsilon_{eff})^2/12\pi$, whereas $Z_2$ can dominantly decay directly to pairs of DM fermions,  $\Gamma_2(DM) /m_2={\rm PS}\cdot (g_DT)^2/12\pi$, where 
`PS'  is a simple phase space factor, \ie,  PS $=(1-4m_\chi^2/m_2^2)^{1/2} (1+2m_\chi^2/m_2^2)$. $Z_2$ can also decay, like $Z_1$,  
into SM fields but with a partial width that also is highly suppressed, \ie,  
$\Gamma_2(SM)/m_2=(e\epsilon_{eff}T)^2/12\pi$, which can generally be neglected but will be included here for completeness since we will sometimes approach the kinematic region 
where PS $\to 0$. As noted above, this has important implications for accelerator searches for the DP. For numerical purposes we can conveniently express this DM annihilation rate in units 
of $\sigma_0 = 10^{-26}$ cm$^3$s$^{-1}$ which sets the typical scale for that required to obtain the observed relic density (recalling that the required Dirac fermion annihilation rate to 
achieve this density for DM masses in this mass range of interest  is $\simeq 4.5\sigma_0$\cite{Steigman:2015hda,Saikawa:2020swg}) as 
\begin{equation}
\frac{\sigma \beta_{rel}}{\sigma_0}\equiv\frac{g_D^2}{e^2} ~\Big(\frac{\epsilon_{eff}}{10^{-4}}\Big)^2~ \Big(\frac{100~\rm{MeV}}{m_1}\Big)^2 ~\frac {\sigma}{\sigma_0}.
\end{equation}
\begin{figure}[htbp]
\centerline{\includegraphics[width=5.5in,angle=0]{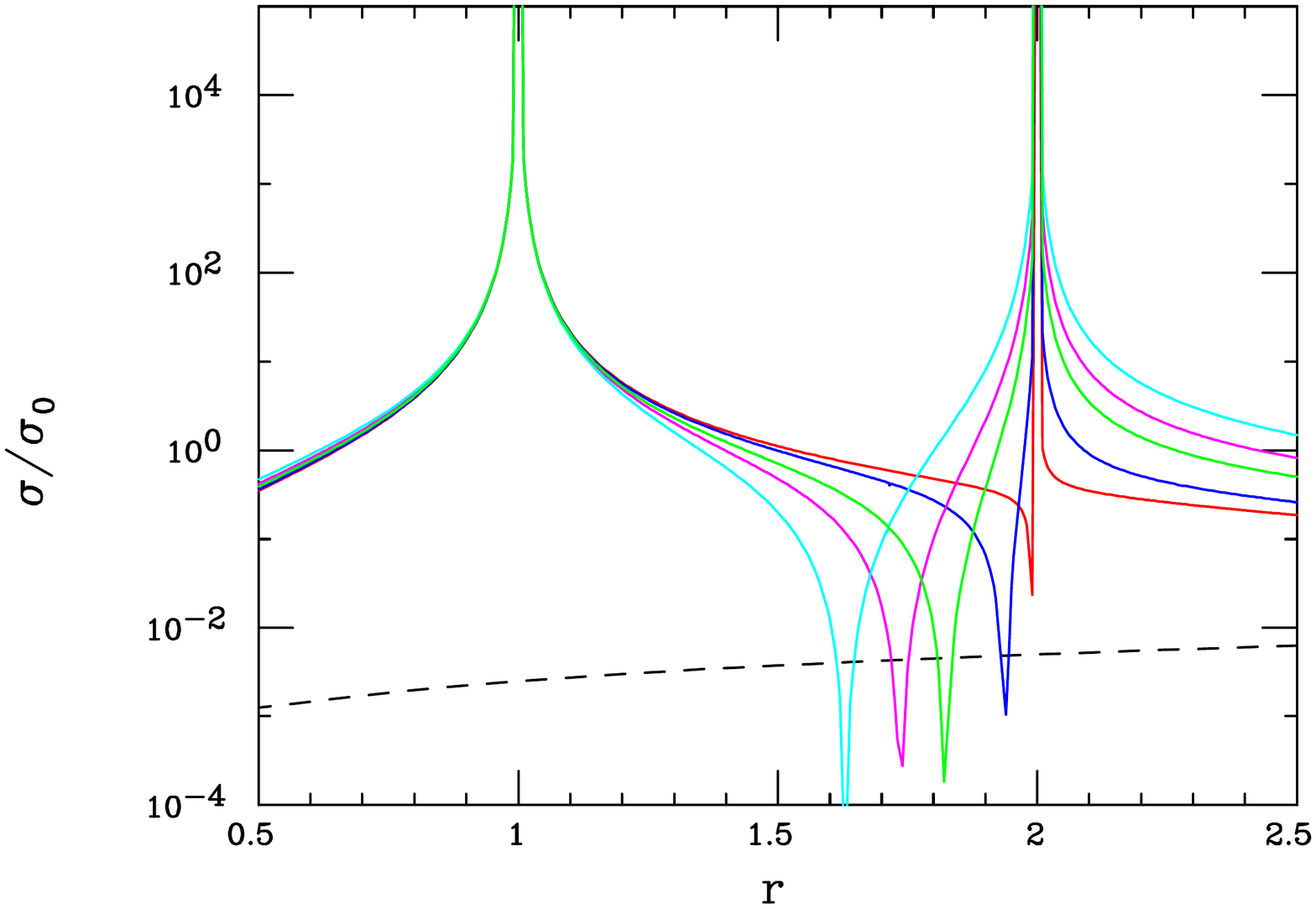}}
\vspace*{-2.3cm}
\centerline{\includegraphics[width=5.5in,angle=0]{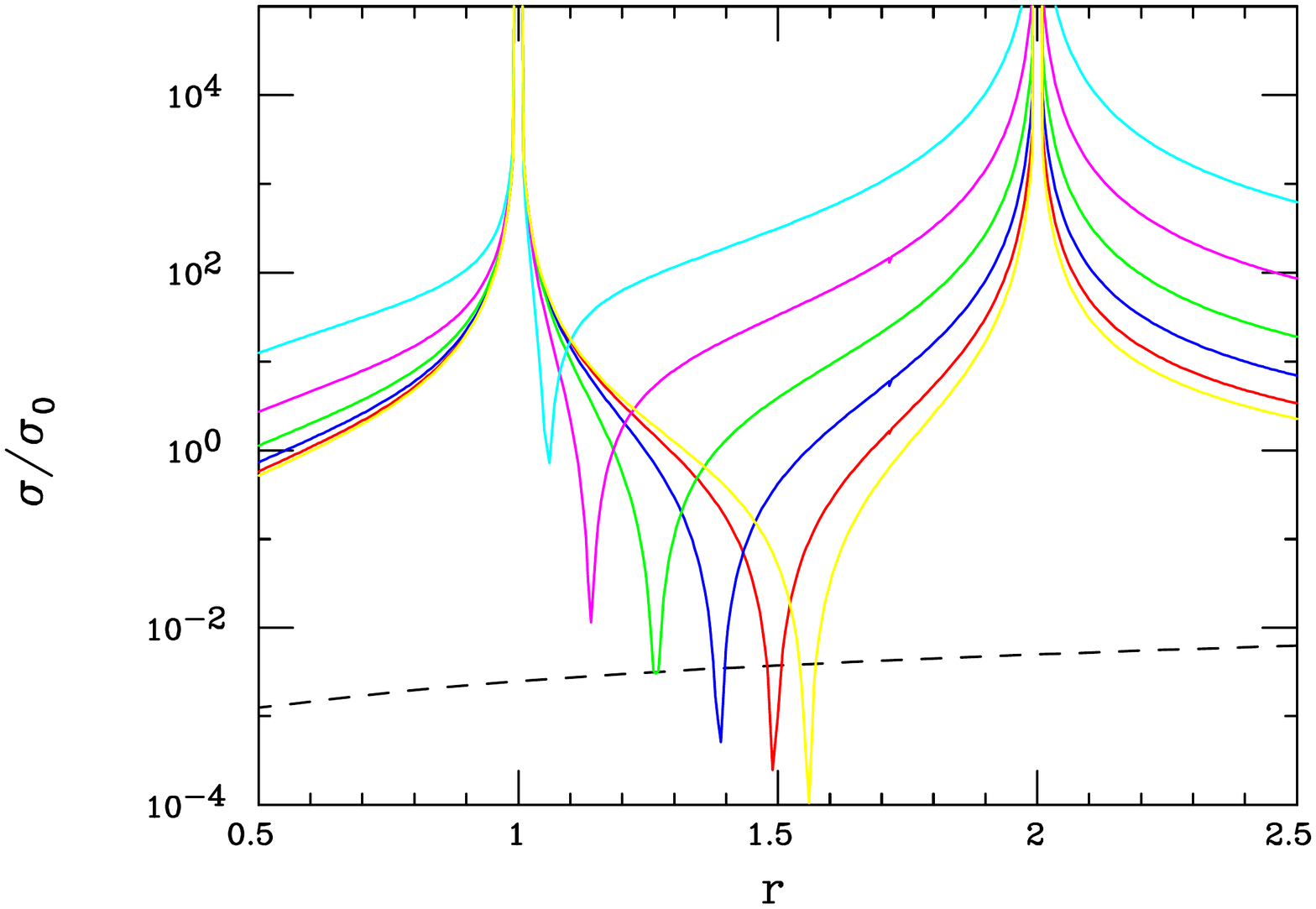}}
\vspace*{-1.30cm}
\caption{Dirac fermion DM annihilation cross section in the $\beta_{rel}^2\to 0$ limit as described in the text assuming that $m_1=100$ MeV, $\epsilon_{eff}=10^{-4}$ and $g_D/e=1$ in units of 
$\sigma_0=10^{-26}$ cm$^3$s$^{-1}$, shown as a function of $r=2m_{DM}/m_1$. Here it is also assumed that $m_2/m_1=2$ and also that, for the minimum, from 
right to left, (Top) $T=0.1, 0.3,0.54 (\simeq \tan \theta_w),0.7,0.9$ and (Bottom) $T=1.05,1.2,1.5,2,3,5$, respectively. In both panels, the dashed line represents the approximate 
upper bound on this cross section allowed by the CMB. }
\label{t0}
\end{figure}
\begin{figure}[htbp]
\centerline{\includegraphics[width=5.0in,angle=0]{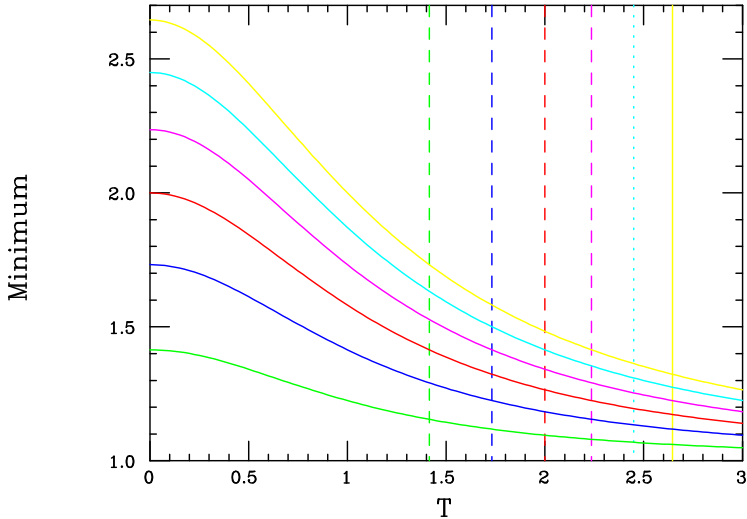}}
\vspace*{-1.50cm}
\caption{Location of the $\beta_{rel}^2\to 0$ cross section minimum in units of $m_1$ as a function of $T$ assuming that, from top to bottom, $m_2^2/m_1^2=7,6,5,4,3,2$, respectively. The 
vertical dashed lines show the maximum allowed value of $T$ for, from left to right, the corresponding value $m_2^2/m_1^2=2,3,4,5,6,7$, respectively. In each case, the region to the 
right of the dashed line is unphysical and so excluded.}
\label{minimum}
\end{figure}

As noted above, during the CMB and at present times, temperatures are sufficiently low so that taking $\beta_{rel}^2,\beta_\chi^2\to 0$ becomes an 
excellent approximation and thus we can assume that $s=4m_{DM}^2$ in such circumstances.  Consider the sample case with the parameter choices $m_2/m_1=2$ with 
$g_D/e=1(0.1)$, $m_1=100$ MeV and $\epsilon_{eff}(<\epsilon)=10^{-4}$ which we will typically employ as basic realizations of our setup. These choices are consistent with the 
present searches for DP production in both the visible as well as invisible decay channels\cite{Fabbrichesi:2020wbt}. Since the cross section approximately factorizes 
as seen above, it is straightforward to obtain the corresponding results for any other choices of 
$g_D/e$, $m_1$ and $\epsilon_{eff}$. For such a parameter set we can completely determine the DM annihilation cross section in the low velocity limit as a function of 
$r=2m_{DM}/m_1$ assuming different values of the parameter $T$ as input; the results of this calculation are shown in Fig.~\ref{t0} assuming that $g_D/e=1$ for purposes of demonstration. 
Here we see the presence of the two resonance peaks with a series of destructive minima lying between them; the location of the saddle minimum 
is seen to move closer to the $Z_1$ hump as the value of $T$ increases as expected from the discussion above. However, we {\it cannot} continually push this minimum to lower values 
of $r$ since $T$ has a maximum value, \ie, $T^{max}=\lambda_R=2$ in the present case, and thus the two furthest left curves in the lower panel are {\it not} actually allowed by this constraint 
and appear here only for the sake of comparison. We note that the range of parameters comfortably satisfying this CMB constraint is rather modest (to say the least) when 
$g_D/e=1$ is assumed. 

To further clarify these points, Fig.~\ref{minimum} shows the location of the $\beta_{rel}^2\to 0$ annihilation cross section minimum as a function of $T$ for various values of the $Z_{1,2}$ 
mass ratio, $m_2/m_1$; also shown is the corresponding upper bound on $T$ for the same range of values of $m_2/m_1$ that we have determined previously above. For a fixed 
$T$ the location 
of the minimum will move to larger (smaller) values as $m_2/m_1$ increases (decreases) and similarly, for fixed $m_2/m_1$ the value of the minimum location will decrease (increase) as 
$T$ increases (decreases).  However, we see that due to the bound on $T$ from above, the location of the allowed {\it physical} minimum can never be pushed to a value of $r$ 
smaller than that given by
\begin{equation}
r_{min}=\Big(\frac{\sqrt s_0}{m_1}\Big)_{min} =\Big[\frac{2\lambda_R^2}{1+\lambda_R^2}\Big]^{1/2}\,,
\end{equation}
for a given $\lambda_R$ so that, \eg, for $\lambda_R=m_2/m_1=2(3)=T^{max}$, $r_{min}=\sqrt {1.6}(\sqrt {1.8})$ and this minimum asymptotes to the value 
$\sqrt 2$ as $\lambda_R\to \infty$. 

Returning now to Fig.~\ref{t0}, we see that, quite generally, the suppressed saddle region between the resonance humps 
can very easily lead to cross sections of order $\sim$ a few $10^{-2}\sigma_0$ or larger when we choose 
$g_D/e=1$ over a modest mass range given the proper choices of $T$. However, we recall that in the units introduced here the CMB cross section bound is roughly given 
by\cite{Cang:2020exa}  $\sigma_{CMB}/\sigma_0<2.5\times 10^{-3} ~r$ as also can be seen in this Figure. To increase the size of our `zone of comfort' where we quite safely satisfy this 
constraint in the saddle region and for later 
phenomenological reasons, we will chose to shift our default value of $g_D/e$ downward, \ie, to $g_D/e=0.1$, so that all of the model predictions displayed in this Figure will also 
shift downwards by a factor of 100. This value shift now 
provides us with a significantly larger region of parameter space safely satisfying the current (and any near future) CMB constraint discussed above for this range of DM masses; we will 
assume this value of $g_D/e(=0.1)$ in the discussion that follows{\footnote {We note that at this point we could have just as easily instead have assumed that $\epsilon_{eff} =10^{-5}$ 
to recover the same reduced cross section as these are both simple overall numerical factors. However, this smaller value of $\epsilon_{eff}$ is somewhat more difficult to arrange at the 
1-loop level and the benefits of the choice of reducing the coupling ratio $g_D/e$ instead will be made more obvious below.}}.  

As previously noted, the successful suppression of the annihilation cross section at CMB times in this set up implies essentially identical annihilation rates today so that DM annihilation 
should not be observable in indirect detection experiments at the canonical  $\sim 10^{-26}$ cm$^3$ s$^{-1}$ rates normally anticipated.

\begin{figure}[htbp]
\centerline{\includegraphics[width=6.0in,angle=0]{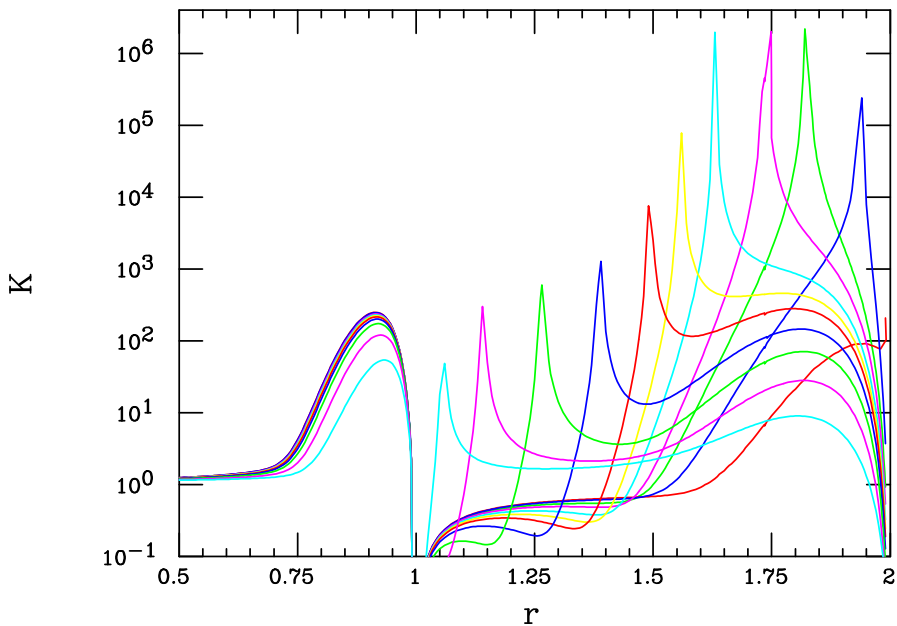}}
\vspace*{-1.50cm}
\caption{The cross section enhancement ratio, $K$, as a function of $r$ assuming $x_F=20$ and $\lambda_R=m_2/m_1=2$. From right to left the curves correspond 
to $T=0.1,0.3,0.54,0.7,0.9,1.05,1.2,1.5,2,3,5$, respectively, as were employed in the previous Figure for comparison.}
\label{Kfact}
\end{figure}

We have now obtained annihilation cross sections easily satisfying the CMB constraint as $\beta_{rel}^2\to 0$ for a respectable large range of parameters. However, we also  
must show that the thermal effects at the time of freeze out can yield a sufficiently large value of $<\sigma \beta_{rel}>_{FO}\simeq 4.5\sigma_0$ for the {\it same} set of input parameters, 
$g_D, \epsilon_{eff}, m_1$. Note that due to the overall parameter factorization exhibited in Eq.(21), the required cross section enhancement factor, $K$, as will be defined below, is 
{\it independent} of the specifically chosen values of $g_D/e$ and $\epsilon_{eff}$ and will instead depend {\it solely} upon the values of the kinematic parameters 
$r, \lambda_R$ and $T$ as well as the temperature at freeze-out, $T_{FO}$.  

At freeze-out, after some algebra, the thermal averaged cross section can be written as (see, \eg, Refs.\cite{Arcadi:2017kky,Plehn:2017fdg}) 
\begin{equation}
<\sigma \beta_{rel}>_{FO}=\frac{8x_F}{K_2^2(x_F)} ~\int_{\gamma_{min}}^\infty ~d\gamma~\gamma^2(\gamma^2-1)K_1(2\gamma x_F) ~\sigma_{\bar \chi \chi \to {\rm SM}}\,,
\end{equation}
where here the role of `SM' will still be played by the $e^+e^-$final state as above,  $x_F=m_\chi/T_{FO} \simeq 20-30$, $K_{1,2}$ are the familiar modified Bessel functions and 
$\gamma=\sqrt s/2m_\chi$ with $\gamma_{min}=1$ here; note that it is only $\sigma$ and {\it not} $\sigma \beta_{rel}$ that appears inside of the integrand in this expression. We now define 
the `enhancement factor', $K$, as the ratio of the annihilation cross section at freeze-out to that obtained during the CMB when $\beta_{rel}^2\to 0$, discussed above, \ie,  
\begin{equation}
K=\frac{<\sigma \beta_{rel}>_{FO}}{<\sigma \beta_{rel}>_{CMB}} \,,
\end{equation}
where we will require, roughly, that $K\sim$ a few $10^3$ or so to get the necessary numerics to work out properly. We gain stress that $K$ itself does not depend on the values of 
$g_D, \epsilon_{eff}$ or even $m_1$ to a rather good approximation since they simply cancel in this ratio but instead depends only upon the two mass ratios and the value of $T$. 
To be specific, let us assume that $\lambda_R=2$ and $x_F=20$; we can then calculate 
$K$ as a function of $r$ for different values of the parameter $T$ as is shown in Fig.~\ref{Kfact} and then search for the regions where $K$ has the desired range of values. Here we see 
that for roughly the range $0.2 \lsim T\lsim 1.3$, the values of $K$ can easily lie within the desired range of $\sim$ a few $10^3$ or so; this corresponds roughly to the scaled DM mass 
range of $1.40 \lsim r \lsim 1.95$.  For larger values of $T$, the locations of the 
$\beta_{rel}^2\to 0$ cross section minima discussed above are just too far away from the $Z_2$ resonance hump to obtain an adequate enhancement -- especially so if we must also 
require that $T\leq 2$ lies within the physically allowed range. We also note that as $T$ increases the width of the $Z_2$ increases, lowering the peak height, also leading to a 
further suppression of the value of $K$, although this is not numerically a very important effect. For smaller values of $T$ outside the above range, the $Z_2$ coupling is simply 
too weak and the proximity of the minimum too close to the $Z_2$ peak 
to provide the cross section boost that is needed. As a further comment on this Figure, we can also see that the values of $K$ obtainable in this setup from 
the usual\cite{Feng:2017drg,Li:2015tka,Bernreuther:2020koj} resonant enhancement mechanism associated with the $Z_1$ is $\sim 100-200$ and is clearly far too small for our 
purposes by a factor of roughly $\sim 10-30$. 

\begin{figure}[htbp]
\centerline{\includegraphics[width=5.5in,angle=0]{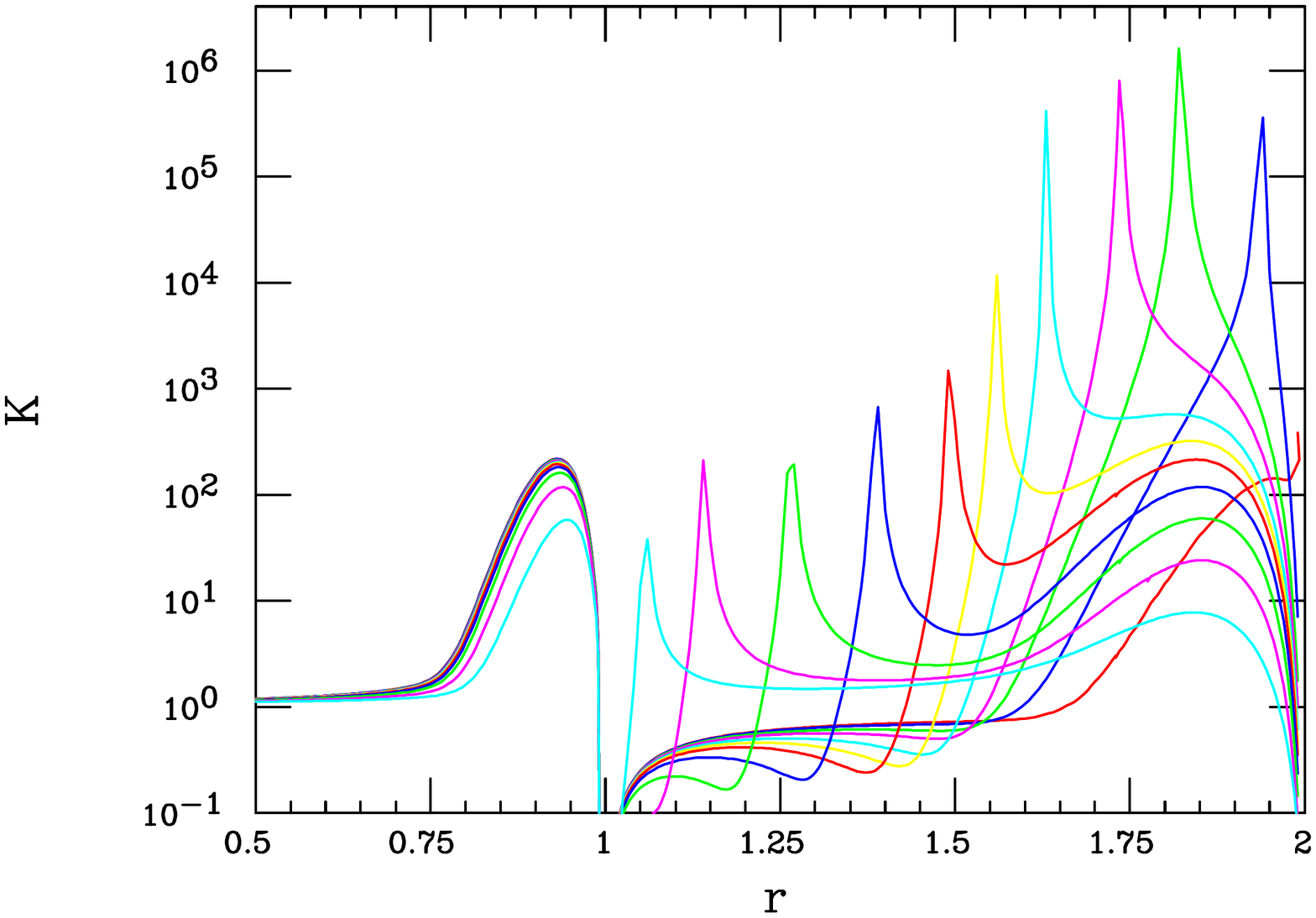}}
\vspace*{-2.3cm}
\centerline{\includegraphics[width=5.5in,angle=0]{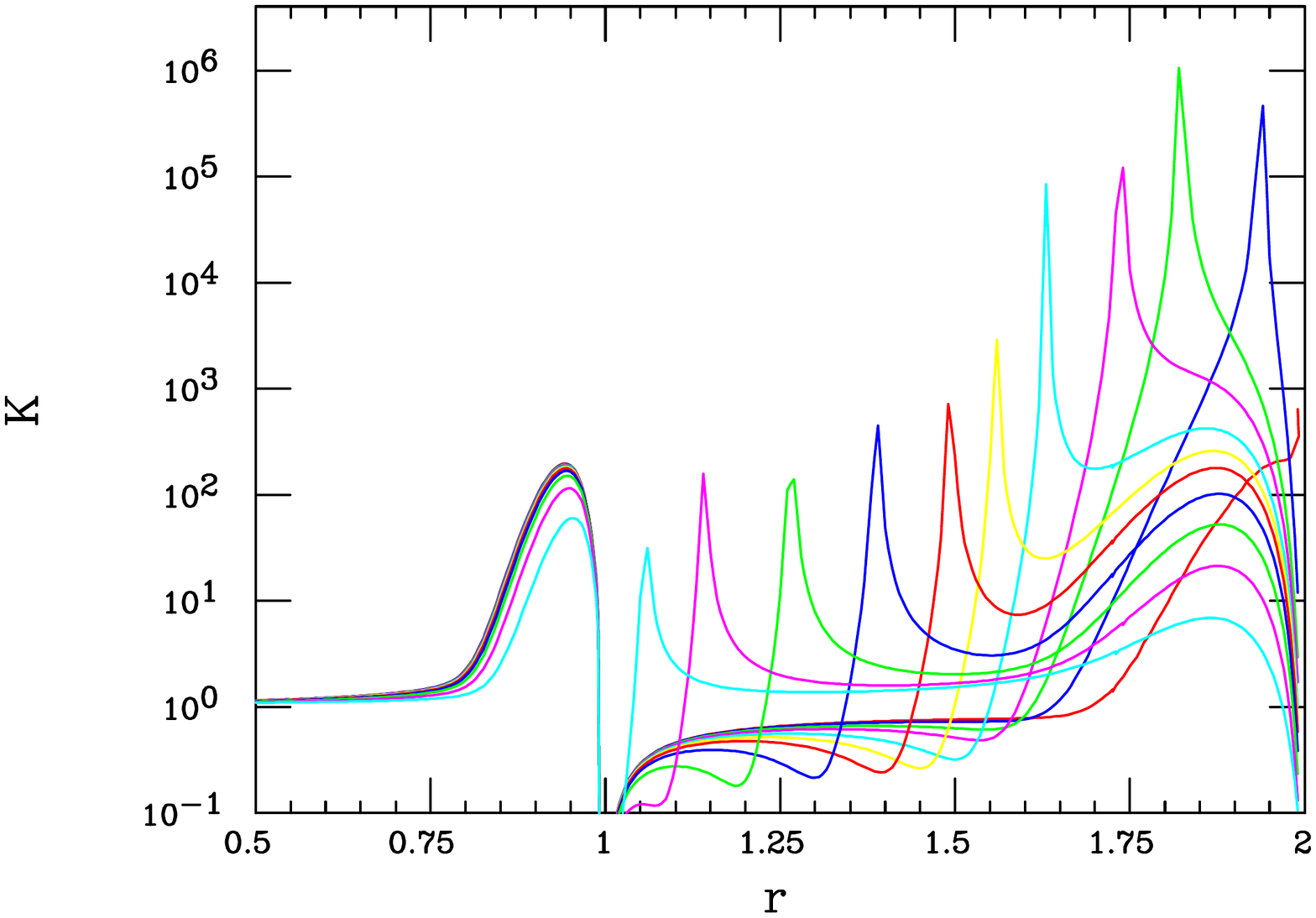}}
\vspace*{-1.30cm}
\caption{Same as the previous Figure but now assuming that (Top) $x_F=25$ or (Bottom) $x_F=30$.}
\label{otherxf}
\end{figure}

It is worthwhile to consider a few variations on this calculation while keeping $\lambda_R=2$ held fixed; we first consider varying out choice of $x_F=20$ to, \eg, larger values, 
\ie, $x_F=25,30$. 
Since $x_F=m_{DM}/T_{FO}$,  an increase in $x_F$ lowers the freeze-out temperature and thus the typical values of $\beta_\chi,\beta_{rel}$ occurring in the DM collision process are 
also reduced since $<\beta_\chi^2> \simeq 2/(3x_F)$ and, hence, so is the typical value of $\sqrt s$. This would imply that for fixed $r$ the DM is less able to feel the influence of the 
second resonance hump 
and we thus expect the value of $K$ to decrease with increasing $x_F$. Fig.~\ref{otherxf} shows what happens when we move to the larger values of $x_F=25$ or 30 and we see that our 
expectations are indeed met and that the range of $T$ over which the value of $K$ is sufficiently large to satisfy our requirements is indeed reduced, but not by a very serious amount. 
For example, even when $x_F=30$, we see that the parameter range $0.2 \lsim T\lsim 1$ easily provides for adequate values of $K$. 

\begin{figure}[htbp]
\centerline{\includegraphics[width=5.0in,angle=0]{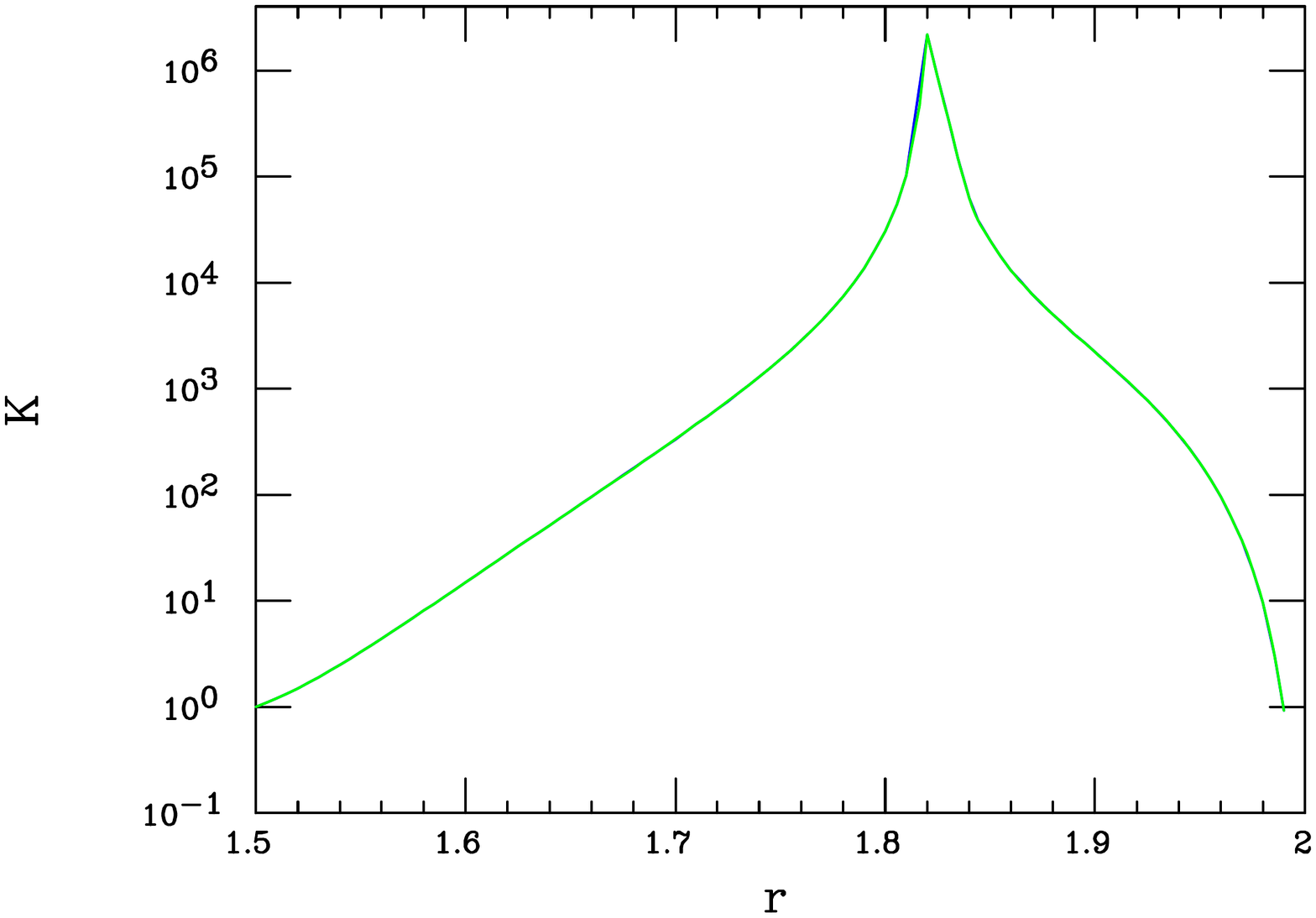}}
\vspace*{-2.3cm}
\centerline{\includegraphics[width=5.0in,angle=0]{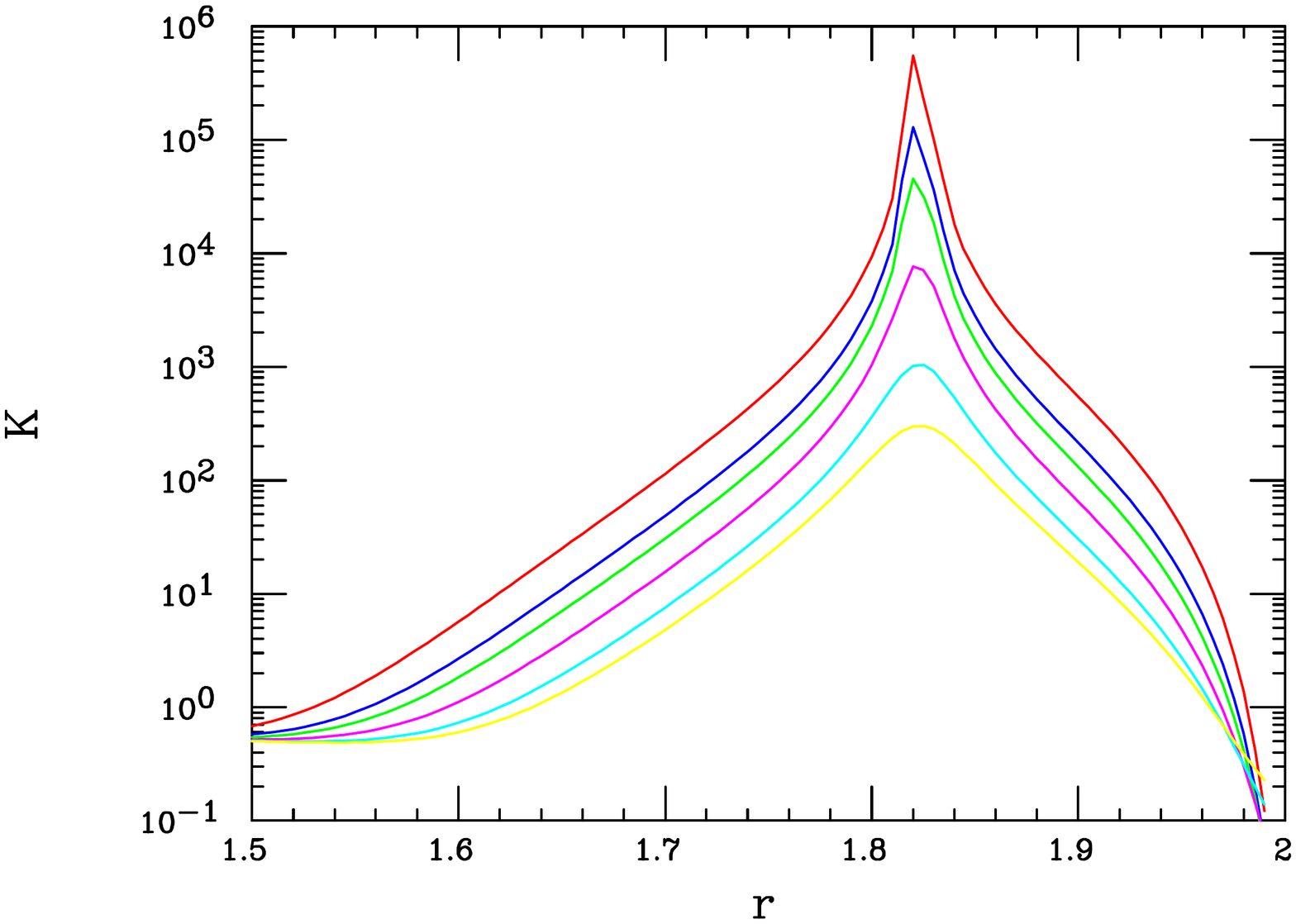}}
\vspace*{-1.30cm}
\caption{Sensitivity of the value of $K$ assuming that $x_F=20$,  $T=0.54$ and $m_2/m_1=2$ (Top) comparing the result obtained employing a running $Z_2$ width (blue) with that 
from fixed width calculation (green) and (Bottom) showing the impact of a larger, fixed $Z_2$ total width assuming that, from top to bottom, 
$\delta \Gamma_2/m_2=0.001,0.003,0.005,0.01,0.02,0.03$, respectively.}
\label{var}
\end{figure}

We briefly consider two other modifications related to the the $Z_2$ total width since its intrinsic `narrowness' as $2m_{DM} \to m_2$ does plays a role in the calculation, specifically, how 
it compares with the thermal `doppler-induced' resonance width. ($i$) One may wonder if the use of `running' decay widths (see, \eg, \cite{Duch:2018ucs}), which scale like 
$\sim \sqrt s$, instead of our default
use of fixed widths might lead to somewhat different results when $x_F,~T$ and $\lambda_R$  (as well as both $g_D/e$ and $\epsilon_{eff}$) are held fixed. The top panel of Fig.~\ref{var} 
addresses this issue for a particular choice of the parameter set; at least in this case we can barely see the difference between the two predictions for $K$ and we conclude that this 
choice likely makes little difference. ($ii$) Since the width of the $Z_2$ becomes $\epsilon_{eff}^2$ suppressed in the limit when $2m_{DM} \to m_2$, one might ask how any additional decays 
of the $Z_2$, into, \eg, other possible dark sector fields, might influence our results due to the increased $Z_2$ width. We recall that in the 
current setup $\Gamma_2/m_2\simeq \Gamma_2(DM) /m_2={\rm PS}~(g_DT)^2/12\pi \simeq 2.4\times 10^{-3}~{\rm PS}~(g_D/e)^2T^2$, where PS is just the phase space factor introduced 
above PS $=(1-4m_\chi^2/m_2^2)^{1/2} (1+2m_\chi^2/m_2^2)$,  
which is generally rather narrow even when $g_D/e=1$. Clearly as this width increases, the height of the $Z_2$ resonance 
hump decreases leading to a suppression of the enhancement of the value of $K$ which is obtainable when all other parameter values are held fixed. {\it A priori}, we don't expect that these 
contributions can be very large since whatever these additional dark fields into which the $Z_2$ can decay may be, they must be heavier than $m_{DM}$ (by definition) so the window for 
their kinematic accessibility is quite small. The lower panel of Fig.~\ref{var} shows the effect of adding these potential {\it ad hoc} contributions to the $Z_2$ width with all of the 
other parameters held fixed. Clearly, if these contributions could become large then there can be a significant reduction in the possible values of $K$ by over an order or magnitude. 
However, as noted, since $m_2$ is not that much larger than $2m_{DM}$ when $\lambda_R=2$, there is very not much of a window for such a large suppression to take place. Of course 
as $\lambda_R$ increases the possibility of such significant contributions can also increase due to the opening up of the allowed phase space. However, as we will see below, such scenarios 
already face other more significant issues.  

So far, we have not spent much time concerning ourselves with the model building constraint ($iv$) above, \ie, that we need to avoid a potentially sizable $s$-wave $\bar \chi \chi \to 2Z_1$ 
process cross section, other than by requiring that $m_\chi=m_{DM}<m_1$ so that, at least when $\beta_{rel}^2\to 0$ during the CMB, this worrisome process is kinematically forbidden. 
However, as is well-known\cite{Griest:1990kh,DAgnolo:2015ujb,Cline:2017tka,Fitzpatrick:2020vba,1837855}, at the time of freeze out, thermal effect can increase the value of 
$\sqrt s$ sufficiently so that this process becomes kinematically allowed although still 
remaining somewhat suppressed by Boltzmann factors. In the current setup, this process occurs through $t-$ and $u-$channel $\chi$ exchange similar to the familiar $e^+e^-$ pair 
annihilation process in QED. Interestingly, if at least part of the fermion DM's mass were to be generated by one or more of the dark Higgs field vevs (which, given our coupling structure, 
is {\it not} the case presently under consideration here and can more easily occur in the case of scalar DM) then those scalars would also contribute to 
this process as $s-$channel exchanges. If we want 
the usual $\bar \chi \chi \to Z_i^* \to e^+e^-$ reaction to remain the dominant DM annihilation process and we don't want the $\bar \chi \chi \to 2Z_1$ process to reduce the amount 
of DM from that we observe, then we must require that the corresponding annihilation 
cross section for the $2Z_1$ final state satisfy the rough bound $<\sigma(2Z_1) \beta_{rel}>_{FO}/\sigma_0  \lsim 1$ ~\footnote{We expect that the $Z_2$ is sufficiently 
massive so that the $Z_1Z_2$ and $2Z_2$ final states do not pose any similar problems.}. To examine this reaction in the present context we make use of the cross section expression 
for this process as given in Ref.\cite{Das:2020rxn} with only a few modifications. This 
reaction is, of course {\it independent} of the values of both $m_2$ and $\epsilon_{eff}$ (which is one reason that it can be so large) but is proportional to $(g_D/e)^4$ 
and will depend on the value of $r$ and, of course, $x_F$, to which we expect some substantial sensitivity since as $x_F\to \infty$ this annihilation rate will vanish due to the Boltzmann 
factors. Recall that the larger the value of $x_F$ the lower the average DM velocity is in the thermal bath and thus the lower is the average value of $\sqrt s$. Based on this Boltzmann 
suppression, semi-quantitatively, we may expect this cross section to to scale roughly as\cite{Griest:1990kh,DAgnolo:2015ujb,Cline:2017tka,Fitzpatrick:2020vba,1837855}  
$\sim e^{[-(m_1-m_\chi)x_F/m_\chi]}=e^{-(2/r-1)x_F}$, which gives a fair approximation to the shape of the numerical results that we obtain below. 

\begin{figure}[htbp]
\centerline{\includegraphics[width=5.5in,angle=0]{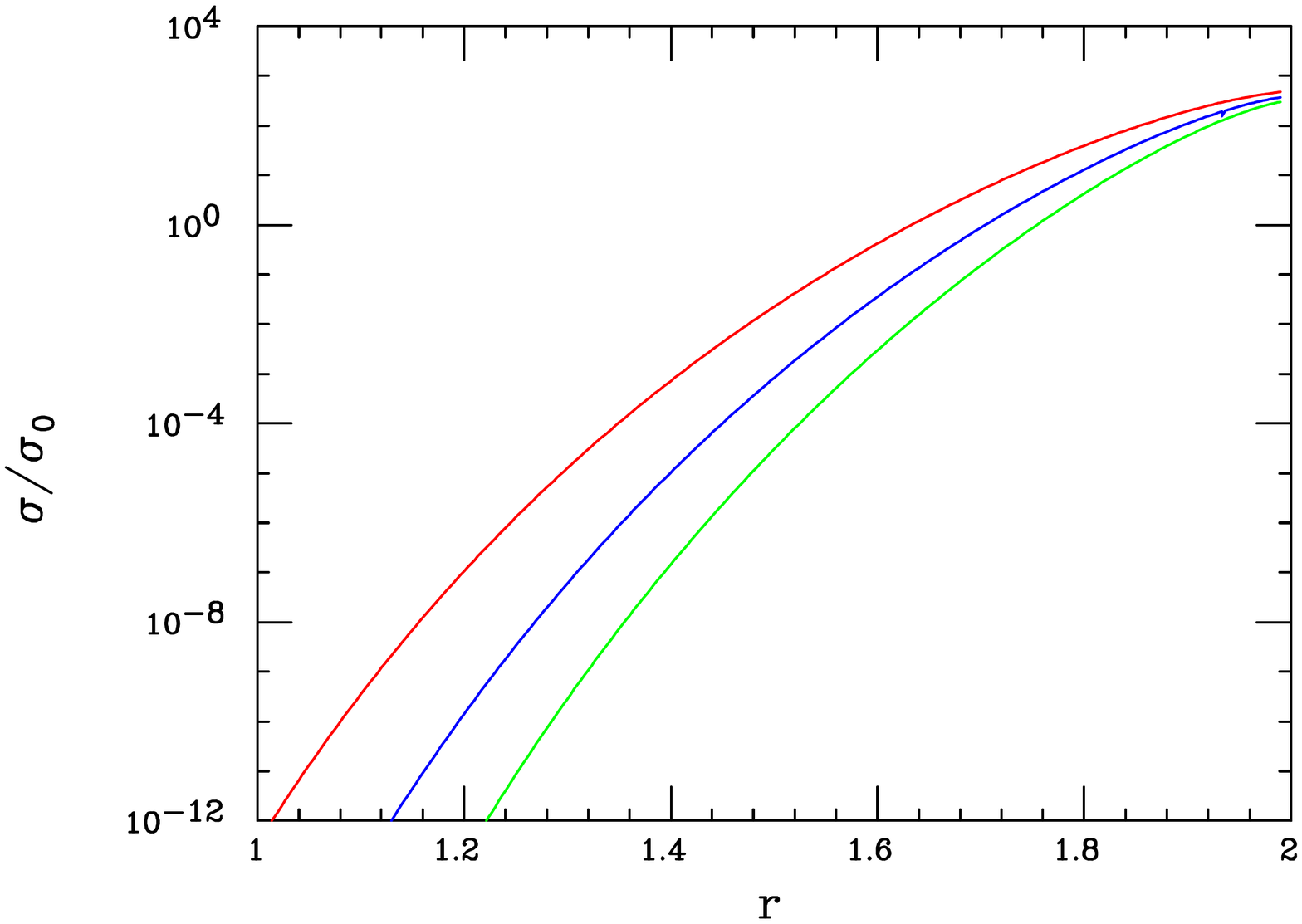}}
\vspace*{-1.50cm}
\caption{Dirac fermion DM pair annihilation cross section into $2Z_1$ via thermal effects, in units of $\sigma_0$, taking $g_D/e=0.1$, as a function of $r$ and assuming, from left to right, 
that $x_F=20,25,30$, respectively.}
\label{vpairs}
\end{figure}

Fig.~\ref{vpairs} shows the result of this cross section calculation as a function of $r$, provided we assume that $g_D/e=0.1$ with $x_F=20,25$ or 30. This result was obtained 
by returning to Eq.(23), adopting the cross section section from Ref.\cite{Das:2020rxn}, as noted above, and now employing $\gamma_{min}=m_1/m_\chi=2/r$ due to the $2Z_1$ 
mass threshold. Here we see several important things: 
($i$) Simply applying the constraint that $<\sigma(2Z_1) \beta_{rel}>_{FO}/\sigma_0  \lsim 1$ implies 
the corresponding rough bounds $r\lsim 1.65(1.71,1.76)$ for $x_F=20(25,30)$. ($ii$) If we 
assume that $\lambda_R=2$ as above, then this constraint tells us that we must require that $0.6-0.7 \lsim T$, depending on the exact value of $x_F$, to avoid this excluded range of 
$r$. Simultaneously,  $T$ is also bounded from above 
if we are to simultaneously obtain a sufficiently large value of $K$ as well as to satisfy the $T\lsim T^{max}=\lambda_R$ limit.  ($iii$) The 
annihilation rate is seen to be is an exponentially strong function of $r$, reflecting the Boltzmann factor, rising extremely rapidly as $r$ increases. For example, we see 
that for values of $r$ only slightly larger than implied by these bounds the annihilation rate is already found to be more than an order of magnitude greater than $\sigma_0$ or possibly larger. 
($iv$) This process is also quite sensitive to $g_D/e$, as noted above, due to its overall $g_D^4$ coupling dependence; this is the main reason for making the choice 
$g_D/e=0.1$ as part of this discussion{\footnote{This choice also renders us safe from the corresponding process where one of the $Z_1$'s is produced off-shell\cite{Rizzo:2020jsm}.}}. 
Changes 
in this parameter will also lead to some substantial modifications on the constraints on the value of $r$ and consequently the value of $T$ as we can see by comparing Fig.~\ref{vpairs} 
and Fig.~\ref{Kfact}. ($v$) Lastly, we note that as $m_\chi/m_1 \to 1$ all of the predictions for the different values of $x_F$ converge to a common result for the cross section. This should be 
no surprise since as $r\to 2$ the amount of additional thermal kinetic energy needed for $\sqrt s$ to exceed $2m_1$ shrinks rapidly to zero and so the cross section 
becomes independent of the temperature. 

\begin{figure}[htbp]
\centerline{\includegraphics[width=5.3in,angle=0]{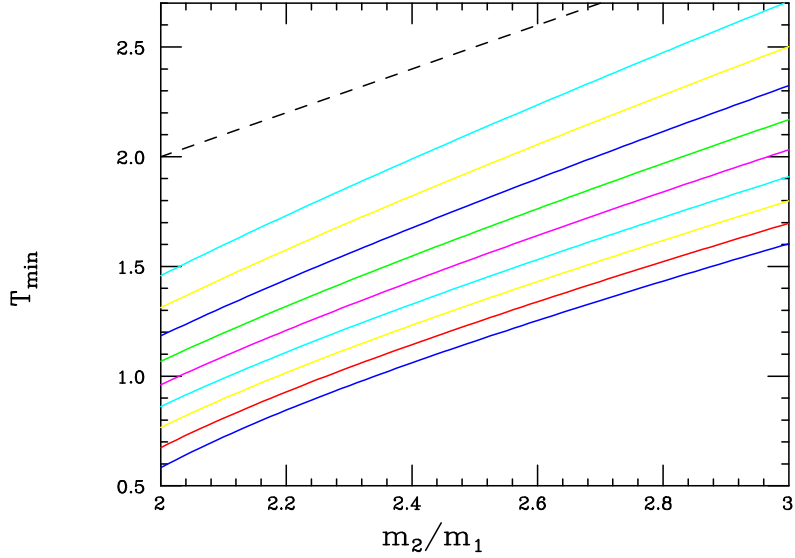}}
\vspace*{-1.50cm}
\caption{Required value of $T=T_{min}$ for the cross section minimum to lie at $\sqrt s_0/m_1=1.4-1.8$, from top to bottom in steps of 0.05, as a function of mass ratio $m_2/m_1$. 
The dashed line represent the maximum allowed value of $T$ as a function of $m_2/m_1$ as described in the text.} 
\label{tmin}
\end{figure}

Given these results, we necessarily must focus on a somewhat narrower model parameter space region. To this end, Fig.~\ref{tmin} displays the value of $T=T_{min}$ for the 
$\beta_{rel}^2\to 0$ annihilation cross section 
minimum to lie at specific values of $\sqrt s_0/m_1$ as a function of the mass ratio $\lambda_R=m_2/m_1$; this is also, very closely, the location 
where $K$ is maximized. Hence, 
for example, if we require a maximum value of $r$ to lie near $r=1.5[1.7]$, which is likely within the most interesting region, then, \eg, for $\lambda_R=2(2.2,2.5,3)$ we will simultaneously 
require that $T$ take on values close to $1.18(1.44,1.79,2.32)[0.77(1.02,1.33,1.80)]$. Of course, this does not guarantee that the value of $K$ which results will be sufficiently large so as 
to meet our needs and to determine that we must perform a detailed calculation as we did for the case of $\lambda_R=2$ above. Note that when this constraint from $2Z_1$ 
production is included only the approximate range $1.40 \lsim r \lsim 1.70$ can now yield a sufficiently large value of $K$ when we assume $\lambda_R=2$.

\begin{figure}[htbp]
\centerline{\includegraphics[width=4.3in,angle=0]{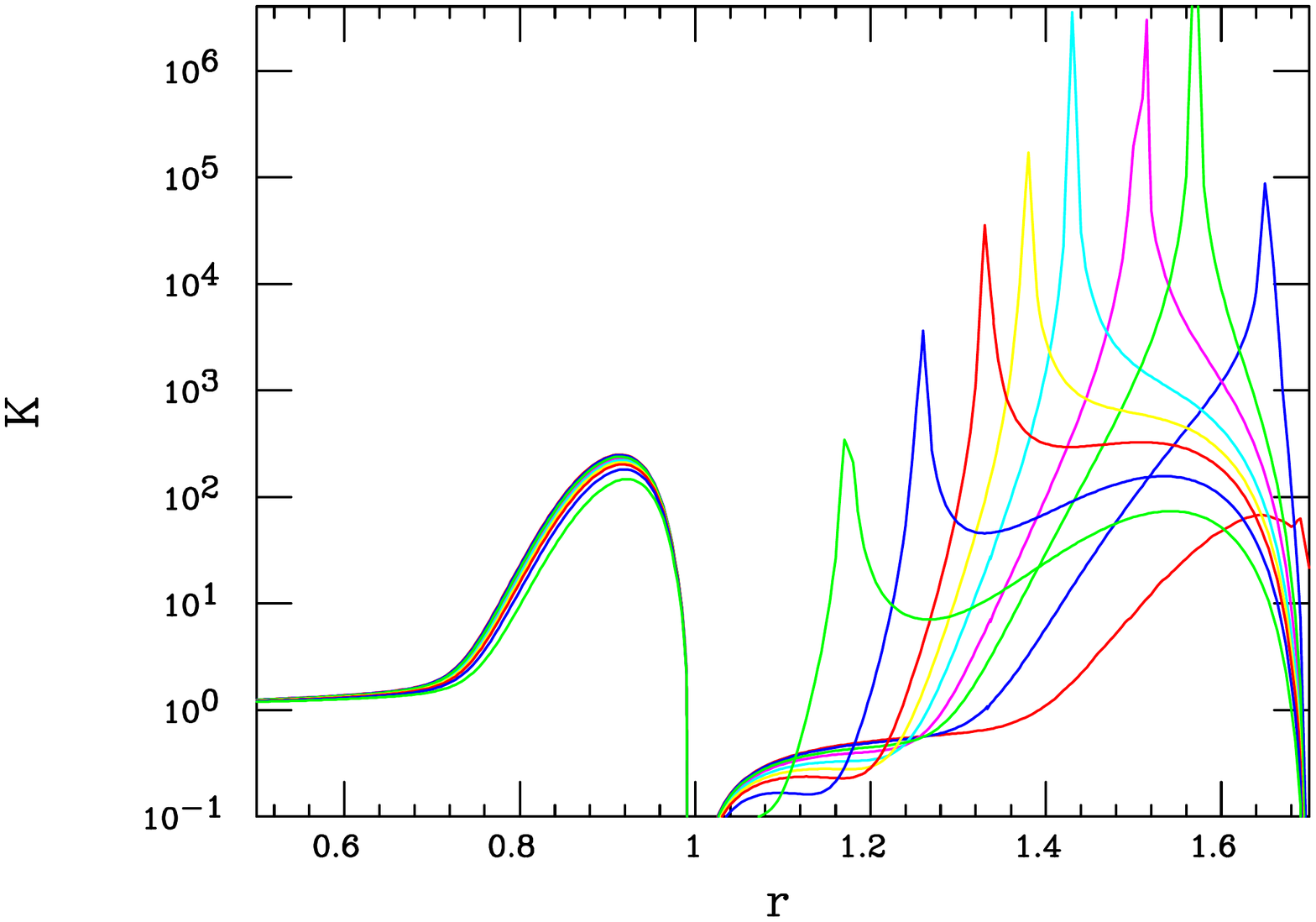}
\hspace {-1.9cm}
\includegraphics[width=4.3in,angle=0]{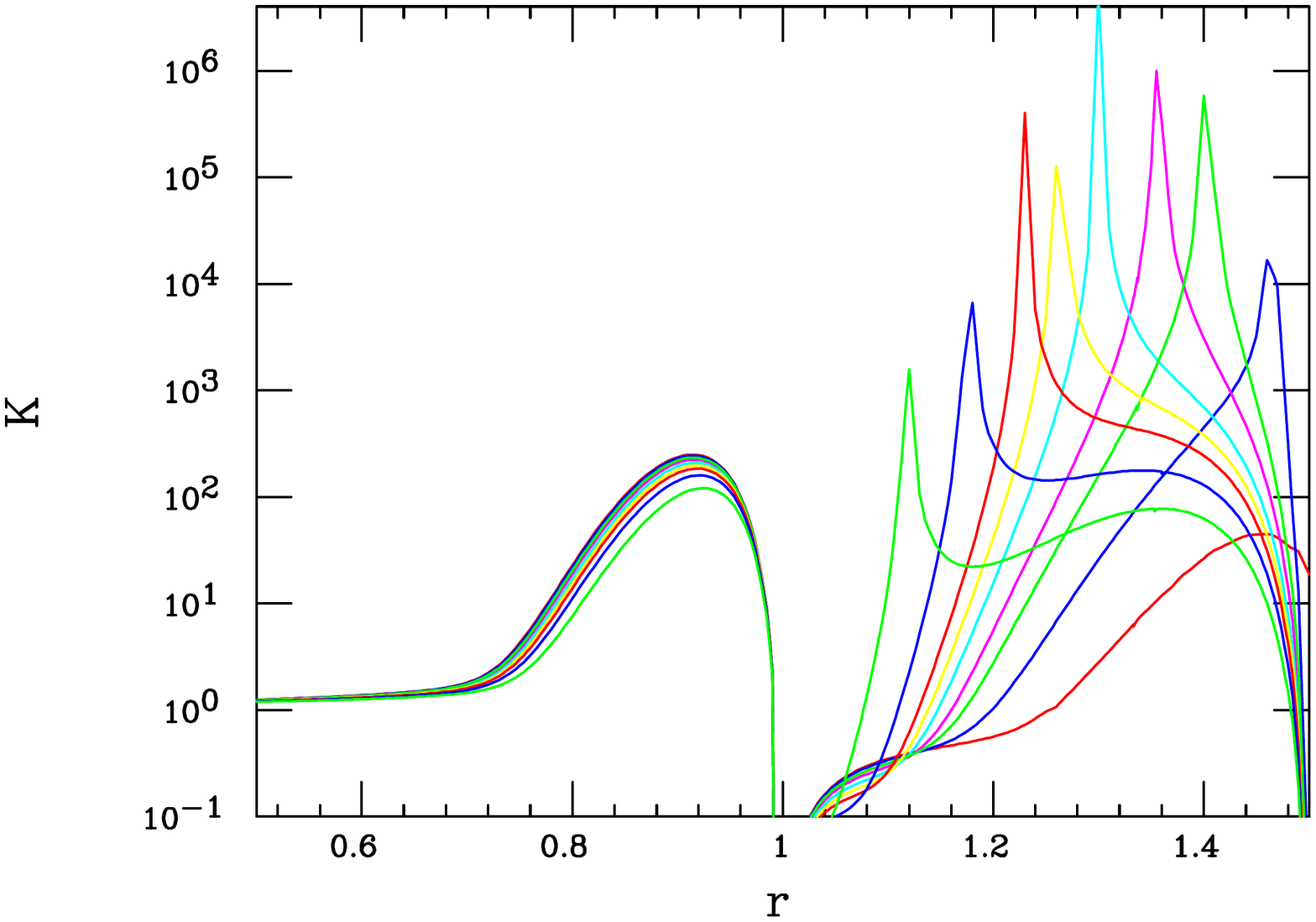}}
\vspace*{-0.6cm}
\centerline{\includegraphics[width=4.3in,angle=0]{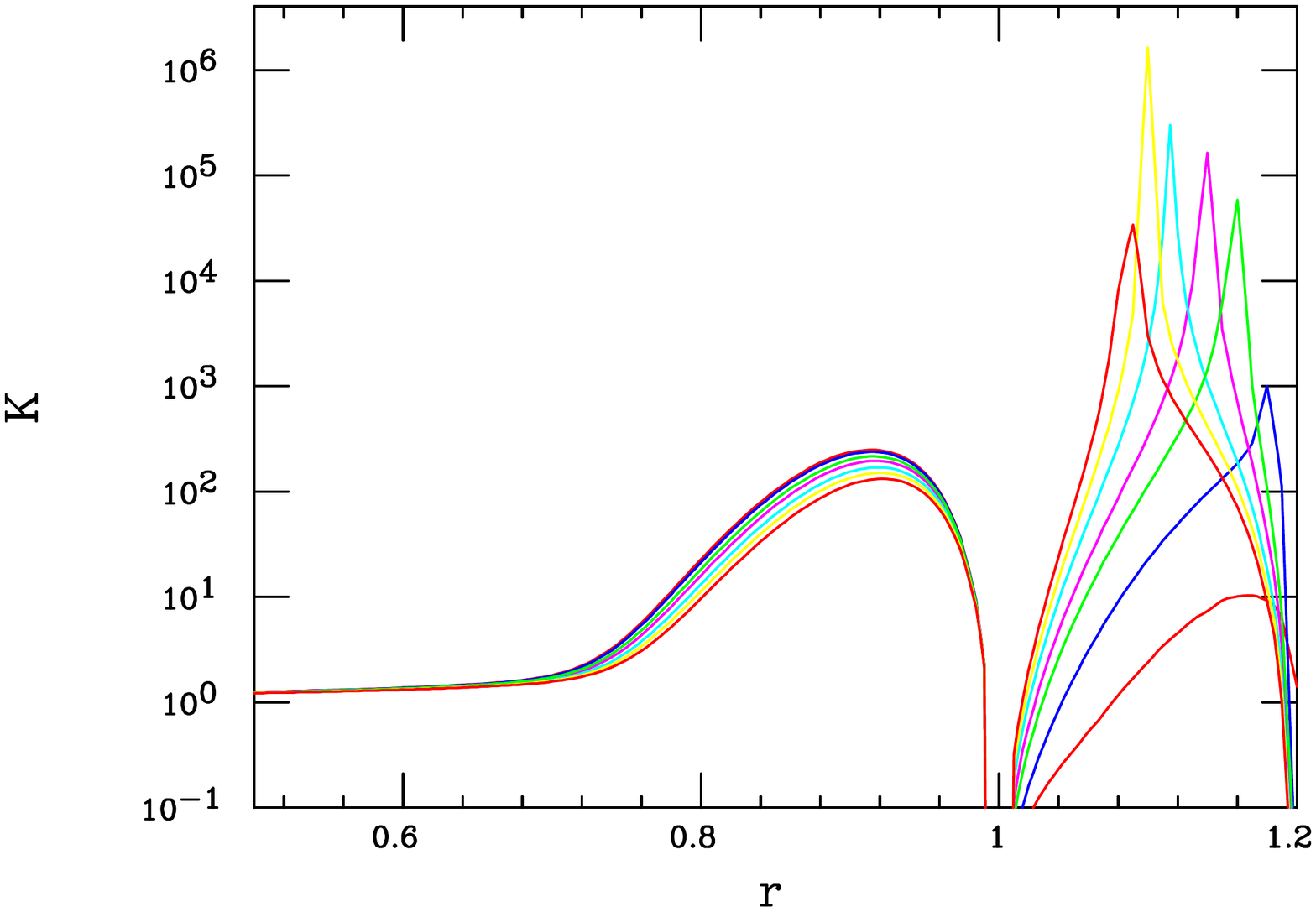}}
\vspace*{-0.70cm}
\caption{Same as in Figure~\ref{Kfact} but now assuming that (Top left) $\lambda_R=1.7$, (Top right) $\lambda_R=1.5$ and (Bottom) $\lambda_R=1.2$, respectively, but now with reduced 
ranges of $T$, still beginning with $T=0.1$ on the right-hand side of each panel as before.}
\label{Kless}
\end{figure}

Due to the non-abelian structure of our setup, there is a second, similarly kinematically forbidden process that we may also be concerned about, \ie, 
$\bar \chi \chi \to Z_{1,2}^* \to W_I^+W_I^-$ where the $Z_{1,2}$ exchanges in the $s$-channel are found to destructively interfere to maintain tree-level unitarity.  The corresponding 
$t,u-$channel exchanges, familiar from the SM, are absent here as the DM, $\chi$, is an $SU(2)_I$ singlet state. One finds, however, that it is always true for the set of parameters 
considered in the present analysis that roughly $(1.1-1.3)m_1\lsim m_{W_I}$. (This further implies that the decay channel $Z_2\to W_I^+W_I^-$ for the on-shell final state will 
open up once $(2.2-2.6)m_1 \lsim m_2$.)   Thus this kinematic suppression coupled with the destructive interference of the two amplitudes in the $s-$channel as well as the absence of 
$t,u-$channel exchanges renders this process far less important than the $2Z_1$ final state we have already considered above when obtaining parameter constraints. This result 
remains true even if other values of $\lambda_R \neq 2$ are considered, a subject to which we now turn.

\begin{figure}[htbp]
\centerline{\includegraphics[width=5.0in,angle=0]{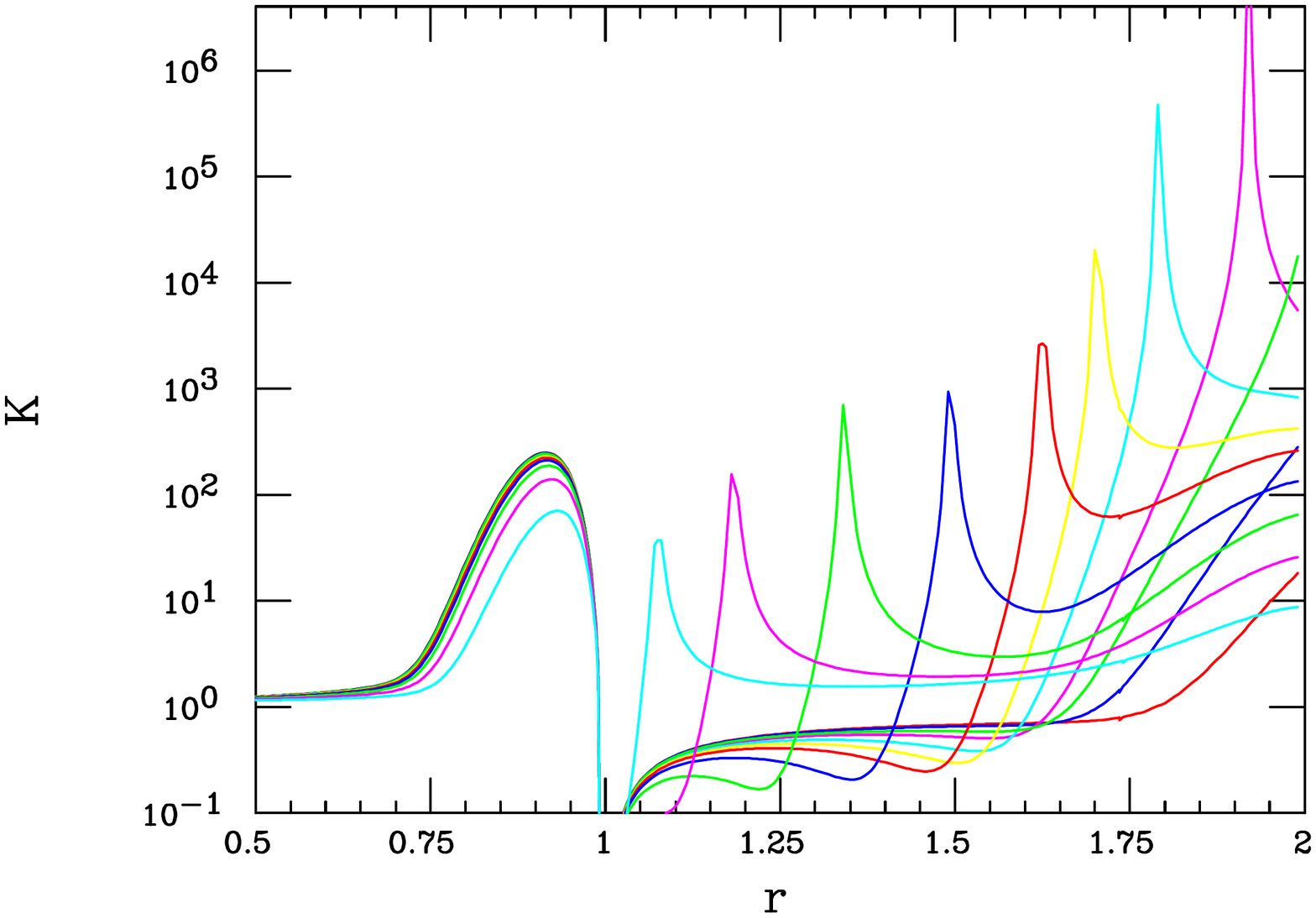}}
\vspace*{-2.3cm}
\centerline{\includegraphics[width=5.0in,angle=0]{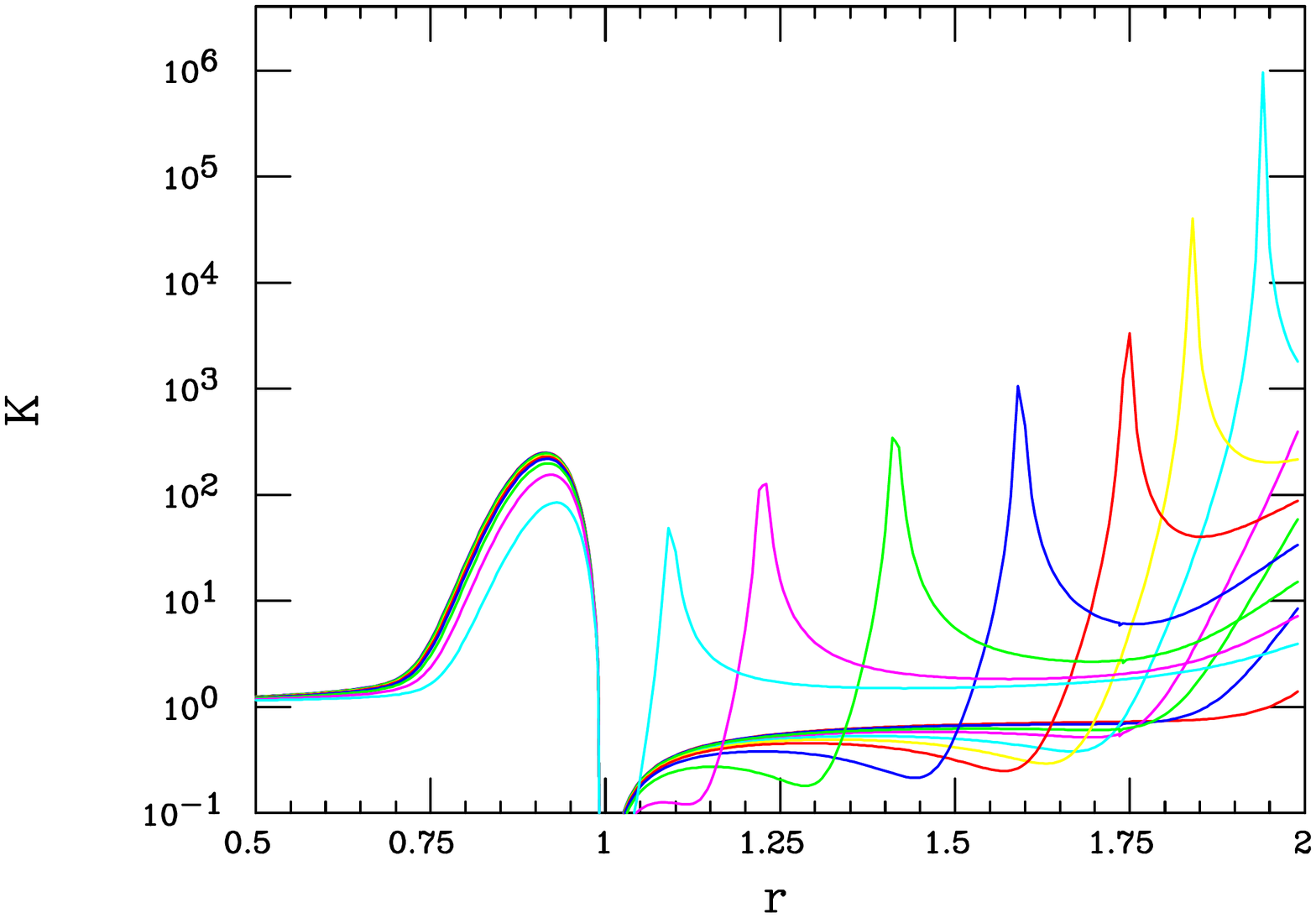}}
\vspace*{-1.30cm}
\caption{Same color codings as in Fig.~\ref{Kfact} but now assuming that $\lambda_R^2=m_2^2/m_1^2=5(6)$ in the top (bottom) panel.}
\label{heavy1}
\end{figure}

Up to this point we have mostly limited our discussion to the case of $\lambda_R=2$ and it behooves us to now ask what happens to our results when this value is modified. First, let 
us consider the case where $\lambda_R<2$: at first thought, this might be very advantageous since as $\lambda_R$ decreases the location of the destructive minimum must move 
to lower values of $r$ thus making it much easier or even trivial to avoid the constraint $r\lsim 1.7$ arising from the $2Z_1$ production process. However, as was already noted 
above in our discussion of Fig.~\ref{Kfact}, it is not advantageous to have $r$ too close to $m_2/m_1$ and lowering $\lambda_R$ significantly decreases the possible range of $r$ 
over which $K$ can be large. Fig.~\ref{Kless} shows the result of our calculation of $K(r)$ as we gradually lower the value of $\lambda_R$ from 1.7 to 1.5 to 1.2 with all the other 
parameters held fixed. When, \eg,  $\lambda_R=1.7$, 
we see that for $0.3 \lsim T \lsim 1.5$, corresponding to roughly $1.25\lsim r \lsim 1.65$, a sufficiently large value of $K$ is obtained while automatically avoiding a large rate for the 
$2Z_1$ DM annihilation mode. However, we see that as $\lambda_R$ further decreases, the allowed range of $T$ is somewhat reduced due to the requirement $T\leq \lambda_R$, but 
that for $r$ is drastically reduced, \ie, $1.15 \lsim r \lsim 1.45$ when $\lambda_R$=1.5 and only the narrow window $1.08\lsim r \lsim 1.17$ when $\lambda_R=1.2$. 
Thus the $\lambda_R<2$ regime remains attractive as long as we do not go to such low values as to highly compress the remaining allowed parameter space.

In the case of increasing $\lambda_R$, our expectation 
is, since we require that both $r\lsim 1.7$ (due to the limit from the $2Z_1$ annihilation cross section) and $T\lsim \lambda_R$ (from model self-consistency), that the values of $K(r)$ will 
be somewhat reduced as $\lambda_R$ increases when all the other parameters are held fixed. The reason for this expectation was noted above: as the $Z_2$ resonance hump moves away 
from the value of $2m_{DM}$, the ability of the the DM to `feel' this resonance sufficiently to increase the annihilation cross section at freeze-out is reduced, hence, leading to a lower value of 
$K$. Clearly, at some point $\lambda_R$ will become sufficient large, with $r\lsim 1.7$, that {\it no} region of the parameter space allowed by other constraints 
produces values of $K$ in excess of the required value of a few $\times 10^3$ and the model again begins to fail. 

\begin{figure}[htbp]
\centerline{\includegraphics[width=4.3in,angle=0]{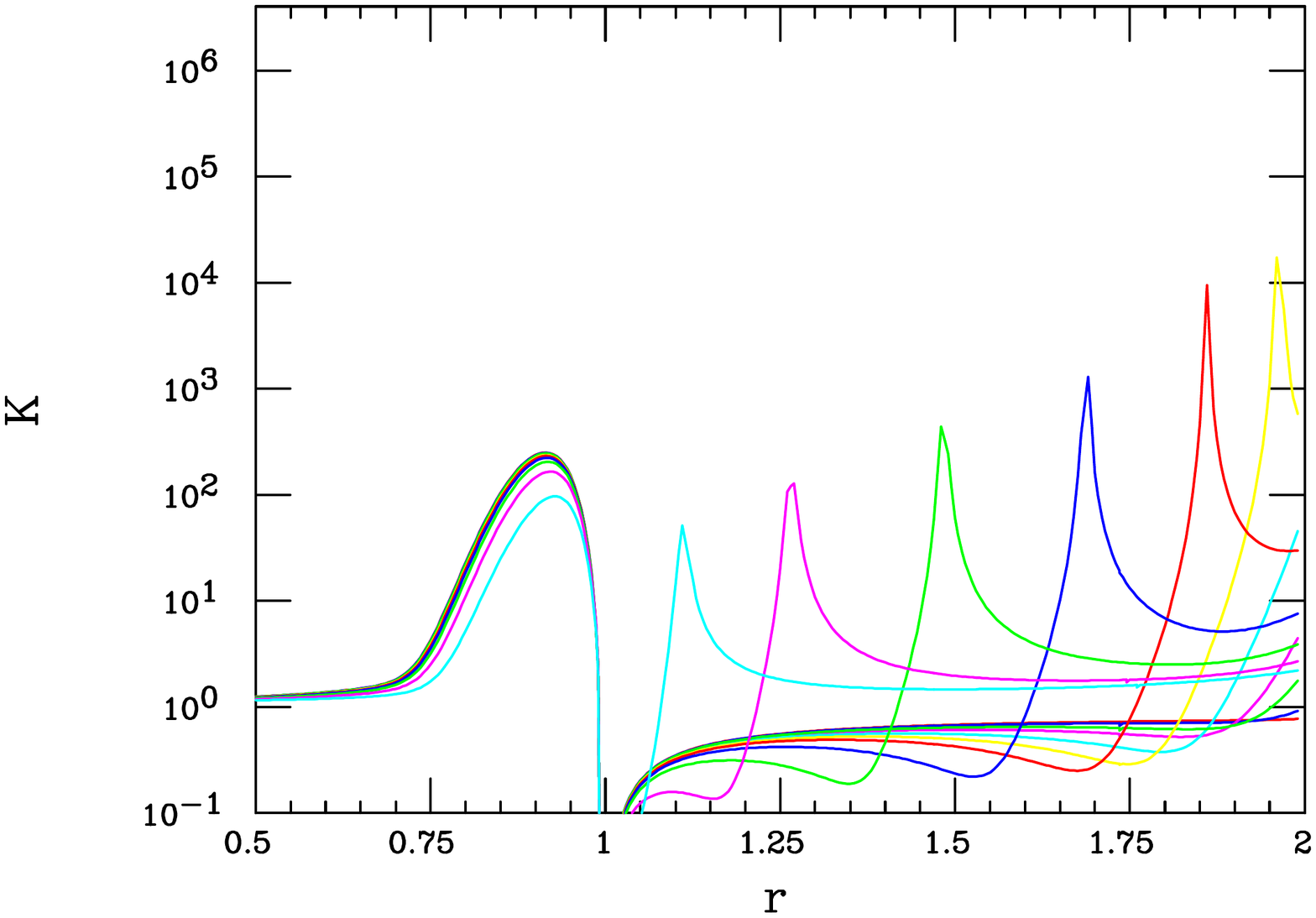}
\hspace {-1.9cm}
\includegraphics[width=4.3in,angle=0]{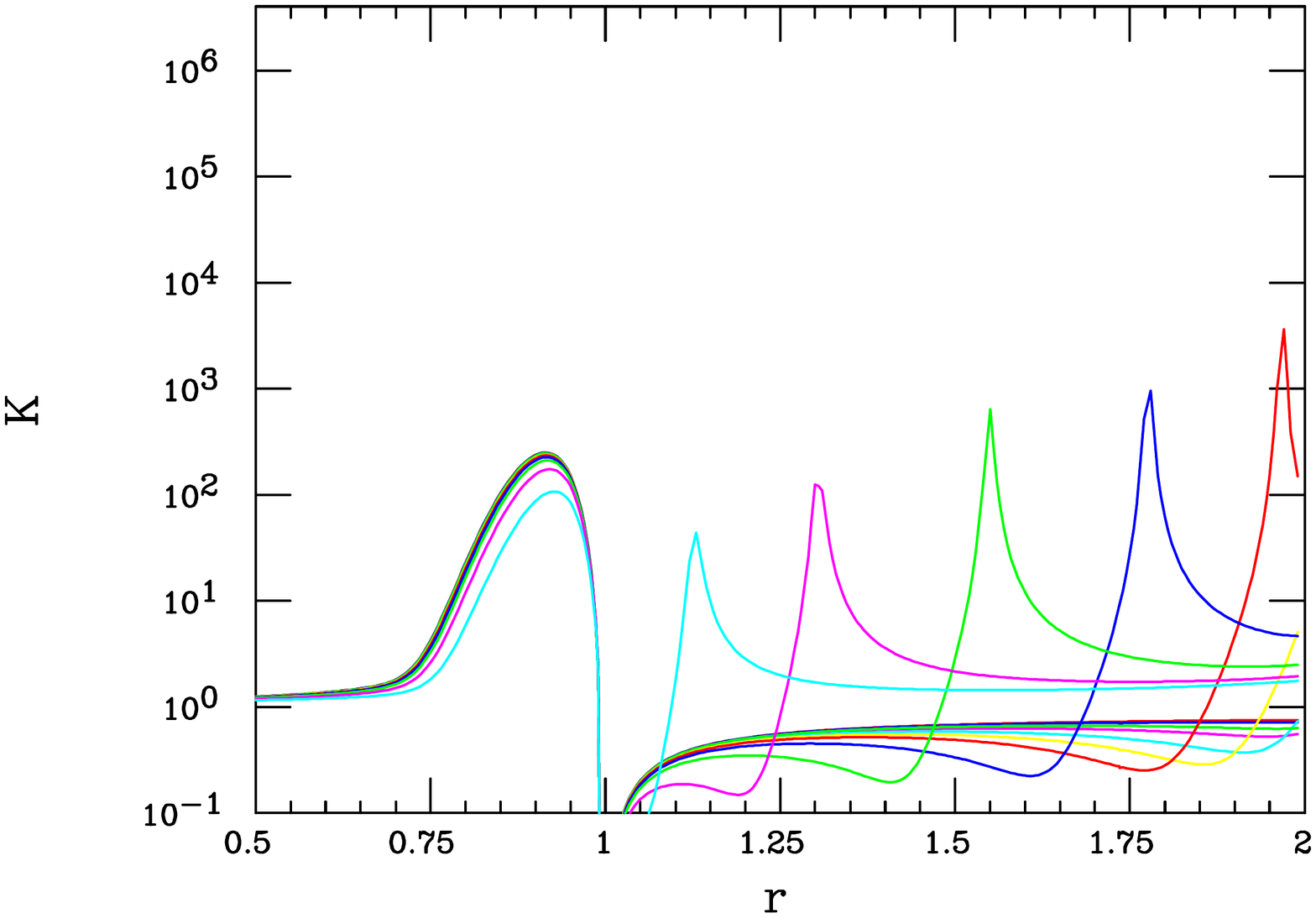}}
\vspace*{-0.6cm}
\centerline{\includegraphics[width=4.3in,angle=0]{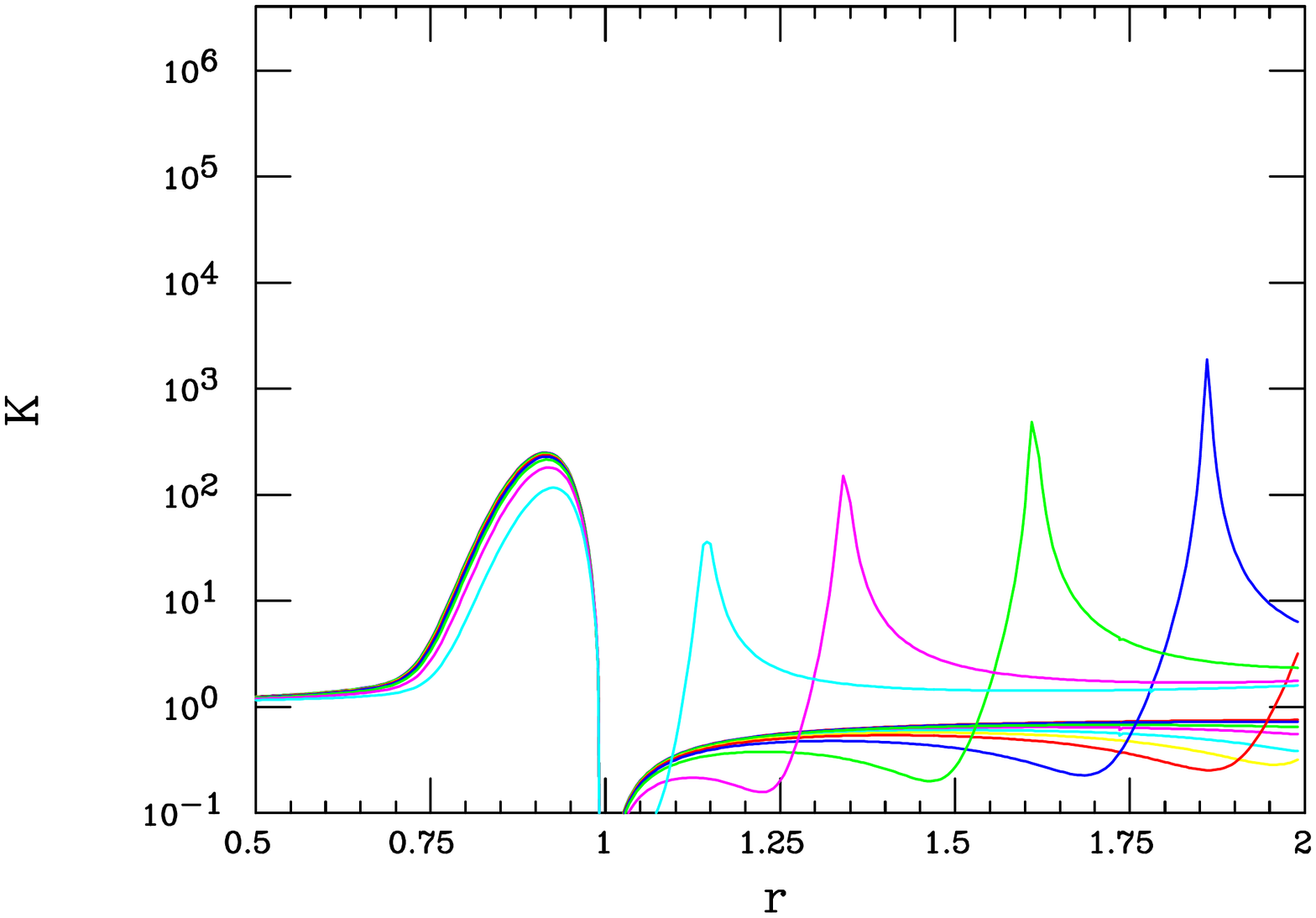}}
\vspace*{-0.70cm}
\caption{Same as the previous Figure but now assuming, from top left to bottom, that $\lambda_R^2=m_2^2/m_1^2=7,8,9$, respectively.}
\label{heavy2}
\end{figure}

Fig.~\ref{heavy1} shows the response of $K$ in our $r$ range of interest to increasing values of $\lambda_R^2$ to 5 and 6 for the same default values of the other parameters as 
considered previously above in Fig.~\ref{Kfact}.  As might be expected, for a fixed value of $r$, the peak of the $K$ distribution moves to higher values of $T$, while for fixed $T$, the peak 
moves to larger values or $r$, more frequently beyond our range on interest as $\lambda_R$ only increases further. Here we see already that for $\lambda_R^2=5(6)$, only the range 
$1.0 \lsim T\lsim 1.2$ corresponding to $1.60\lsim r\lsim 1.70$ ($T\sim 1.20$ with $r\simeq 1.70$) provides a sufficiently large value of $K$ while also avoiding the $2Z_1$ constraint. 
These conflicting requirements are brought home even more strongly in Fig.~\ref{heavy2} where even larger values of $\lambda_R$ are considered. For $\lambda_R^2=7$ we see that 
$T$ is constrained from both directions to lie in a very small region near $\sim 1.3-1.4$ with $r\simeq 1.70$  while for even larger values of $\lambda_R$, no values of 
$K \gsim 10^3$ seem to be obtainable and 
thus no region of parameter space remains tenable given the choices above. From this analysis we see that once $\lambda_R$ becomes much larger than $\sim 2$, the size of the allowed 
parameter ranges rapidly fall to zero essentially forcing us to consider only the range $\lambda_R \lsim 2.6-2.7$. 

\begin{figure}[htbp]
\centerline{\includegraphics[width=5.0in,angle=0]{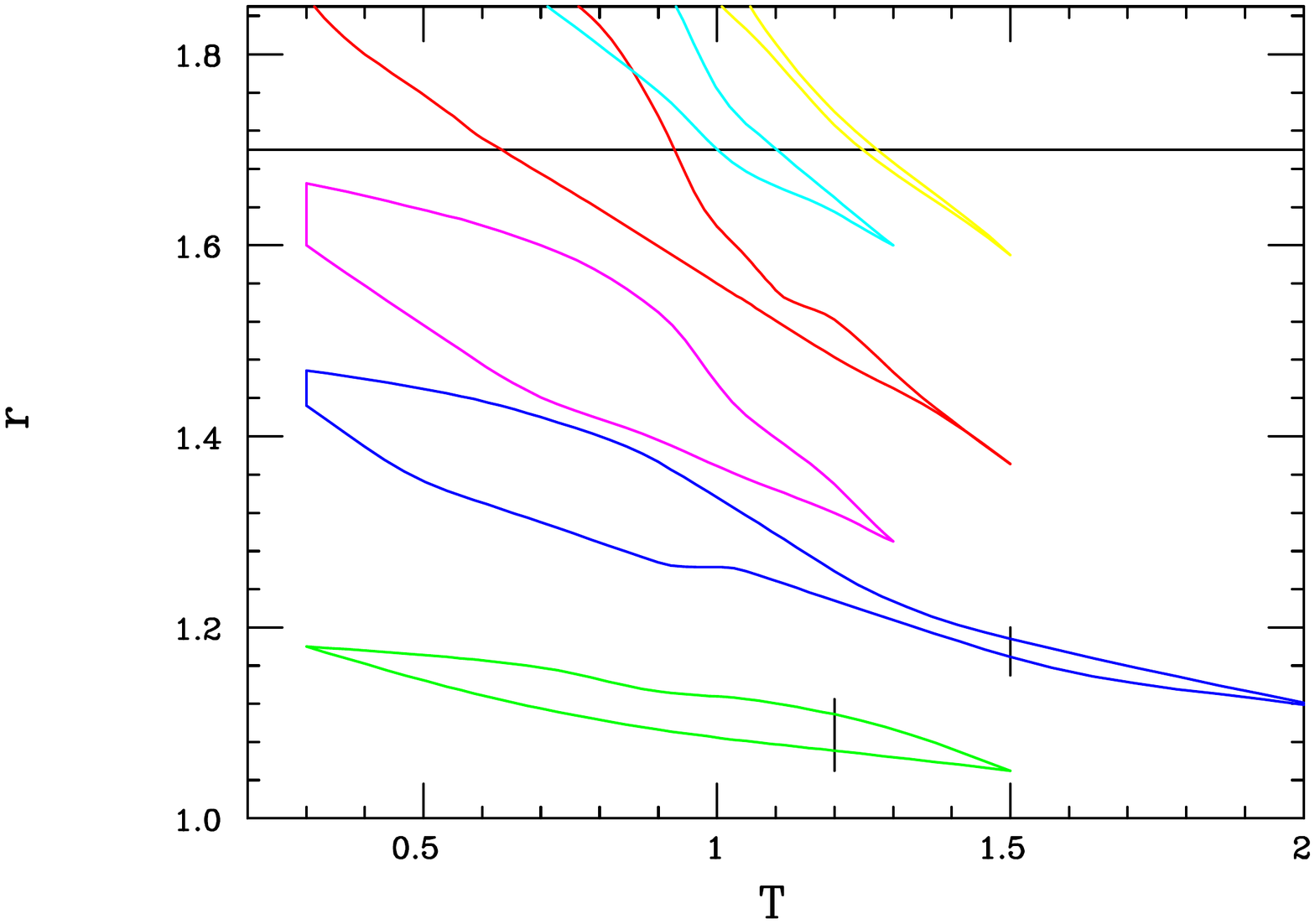}}
\vspace*{-1.50cm}
\caption{Approximate regions in the $T-r$ plane allowed by all of the requirements discussed in the text assuming that $g_D/e=0.1$ and $x_F=20$. From top right to bottom left 
the regions correspond to $\lambda_R=\sqrt 6, \sqrt 5, 2, 1.7, 1.5$ and 1.2, respectively. For larger values of $\lambda_R$, the model points essentially all lie above the `Forbidden DM' 
bound for $2Z_1$ production represented by the horizontal solid line at $r\simeq 1.7$ and are not shown. Note that the lower two regions are cut off on their right sides by the 
$T\leq \lambda_R$ requirement which is shown as the small vertical lines.} 
\label{allowed}
\end{figure}

Fig.~\ref{allowed} provides an overall rough summary of the model allowed parameter space regions as $\lambda_R, T$ and $r$ are all varied assuming that $g_D/e=0.1$ and 
$x_F=20$ are both held fixed. We see that, overall, a significant region of the model parameter space leads to successful results.

Finally, one might wonder what this setup predicts, \eg, for DM direct detection experiments in comparison to the usual single DP scenario due to the existence of the two $Z_i$ exchanges. 
As is well-known, in this DM mass range, $\chi-e$ elastic scattering 
may likely be the most sensitive channel\cite{Aguilar-Arevalo:2019wdi,Aprile:2019xxb,Aprile:2020tmw,Amaral:2020ryn,Arnaud:2020svb,Barak:2020fql}.  
Assuming that the DM mass is in the mass range such that $\mu= m_e m_\chi/(m_e+m_\chi) \to m_e$ and noting that the momentum transfer $Q^2<<m_{1,2}^2$, this cross section is 
given numerically in this scenario by the expression 
\begin{equation}
\sigma_{\chi e} \simeq 2.72 \times 10^{-43}~ {\rm cm^2} ~\Big(\frac{100 ~{\rm MeV}}{m_1} \Big)^4 ~\Big[\frac{g_D/e}{0.1}~\frac{\epsilon_{eff}}{10^{-4}} \Big]^2 ~(1+T^2/\lambda_R^2)^2 \,.
\end{equation}
Note that all of the model dependence that we have been concerned with up to now is quite weak in this case and essentially lies completely isolated within  
the last term appearing here such that, since $T\leq \lambda_R$, we 
must have $(1+T^2/\lambda_R^2)^2 \leq 4$ (at most) in the current setup. Thus, we anticipate at most $O(1)$ changes in this cross section from the predictions of the single DP setup 
with the same input values of $g_D,\epsilon$ and $m_1$. This implies that the indirect detection constraints will only be slightly stronger for these model setups than in the conventional 
single DP scenarios and this strengthening depends directly on the $T/\lambda_R$ ratio.

We also again remind the reader that in the setup described here the lighter $Z_1$ will decay exclusively to visible SM states while the heavier $Z_2$ essentially only decays to 
DM making for  interesting simultaneous signatures in accelerator experiments searching for DP production.

\section{Summary and Conclusions}

The possibility of light dark matter coupling to the SM via the kinetic mixing of a similarly light dark photon with the familiar Standard Model hypercharge gauge boson 
is very attractive for numerous reason, attracting much recent attention. 
Such a scenario can lead to a DM relic density consistent with the Planck measurements via the usual thermal freeze out mechanism in the same parameter range that is accessible 
to multiple future planned experiments. This same accessibility leads to some already significant restrictions on the parameter space of this scenario, if realized in its most simple form, from a 
wide variety of existing experiments. In particular, measurements from the CMB impose rather strong constraints on the DM thermally averaged annihilation cross section at $z\sim 10^3$,  
$<\sigma \beta_{rel}>_{CMB}$, informing us that this quantity must be suppressed by a factor of $K\sim$ a few $10^3$ or so, depending upon the light DM mass, in comparison to the 
analogous cross section at freeze out, 
$<\sigma \beta_{rel}>_{FO}$, that is required to reproduce the observed relic density. This would seem to imply that this reaction must be temperature and/or velocity dependent. Naively, this 
excludes the possibility of DM annihilation being an $s$-wave process as would be the case, \eg, of Dirac fermion DM annihilating via an $s-$channel DP exchange into the SM fermions 
since this type of process is generally temperature/velocity independent. This observation lends support to the possibilities of co-annihilating Majorana DM, which is an $s-$wave but is
Boltzmann suppressed, or $p-$wave annihilating complex scalar DM, which is velocity-squared suppressed, during the CMB epoch. 

In this paper, we have fully examined a previously proposed mechanism by which the $s$-wave Dirac DM annihilation process can be made simultaneously 
consistent with both the relic density and CMB constraints, albeit within a restricted kinematic range. Semi-quantitatively, this requires the existence of (at least) two dark gauge 
bosons, $Z_{1,2}$, by which the DM can pair annihilate via $s$-channel exchange to SM fields -- as noted, this being an $s-$wave process. The $Z_i$ couplings must be such that 
their contributions to this annihilation process destructively interfere, in a manner which is 
independent of the particular SM final state, when the DM pair threshold lies between the masses of these two resonances, \ie, $m_1<2m_{DM}<m_2$. Requiring that 
$m_{DM}/m_1\lsim 0.85$, to avoid the $s$-wave, thermally excited DM pair annihilation into $2Z_1$ (which is not suppressed by KM) while also simultaneously keeping $2m_{DM}$ not too 
far below $m_2$, so that a very strong resonant enhancement from the deep destructive minimum can occur, greatly restricts the parameter space of any potential concrete model. 
Being $s$-wave process, the successful suppression of the annihilation rate at the time of the CMB implies a similar suppression today so that DM annihilation at the canonically expected 
rate should not be observed in current indirect search experiments at the level of $\sigma_0$. 

In order to explore the interplaying roles of these rather restrictive requirements, we constructed a non-abelian, SM-like $SU(2)_I\times U(1)_{Y_I}$ dark sector model but one whose gauge 
symmetry is completely broken leading to $Z_{1,2}$ of comparable masses. The structure of the model's couplings automatically lead to the necessary common destructive interference over 
a significant parameter space region when the $m_1<2m_{DM}<m_2$ condition is satisfied for all SM final states. While the DM couplings of $Z_i$ essentially arise from the gauge 
group structure and the DM representation, here chosen to be a $Y_I/2=Q_D=1$, $SU(2)_I$ isosinglet to help insure it is the lightest dark sector state, the corresponding SM couplings to the 
$Z_i$ are both generated via KM and the various mixing angles required to obtain kinetically normalized fields in the mass eigenstate basis. Within this setup it was found that all of the 
constraints could be satisfied for a respectable range of couplings and values of the mass ratios $m_2/m_1$ and $m_{DM}/m_1$ -- but in a correlated manner. In particular, it was found 
that for the ratio of the $Z_{1,2}$ masses roughly in the range $1.2\lsim  \lambda_R=m_2/m_1 \lsim 2.6$ and a DM to $Z_1$ mass ratio in the (correlated) range 
$1.1 \lsim r=2m_{DM}/m_1\lsim 1.7$, all of our requirements can be easily met simultaneously for corresponding values of the ratio of the product of $Z_2$ to $Z_1$ couplings to the SM 
and DM of O(1). This demonstrates not only a proof of principle but also that realistic models with all of the desired properties can be constructed allowing for light Dirac fermion dark matter. 
One  prediction of this class of models is  that dark photon searches with either visible or invisible decays should (eventually) observe a signal as the lighter $Z_1$ state always 
decays only to the SM while the heavier $Z_2$ will dominantly decay to DM.

Light dark matter with a light mediator below the $ \sim1 $ GeV scale continues to be a very exciting possibility; hopefully, experimental signatures for such scenarios 
will be observed in the not too distant future.

%------------------------------------ ACKNOWLEDGEMENTS ---------------------------------------%
\section*{Acknowledgements}
The author would like to particularly thank J.L. Hewett, D. Rueter and G. Wojcik for very valuable discussions related to this work. This work was supported by the Department of 
Energy, Contract DE-AC02-76SF00515.

%------------------------------------------- REFERENCES -------------------------------------------%

%-------------------------------------------------- END --------------------------------------------------%


\begin{thebibliography}{99}


%\cite{Aghanim:2018eyx}
\bibitem{Aghanim:2018eyx} 
  N.~Aghanim {\it et al.} [Planck Collaboration],
  %``Planck 2018 results. VI. Cosmological parameters,''
  arXiv:1807.06209 [astro-ph.CO].
  %%CITATION = ARXIV:1807.06209;%%

%\cite{Arcadi:2017kky}
\bibitem{Arcadi:2017kky} 
 G.~Arcadi, M.~Dutra, P.~Ghosh, M.~Lindner, Y.~Mambrini, M.~Pierre, S.~Profumo and F.~S.~Queiroz, 
%``The waning of the WIMP? A review of models, searches, and constraints,''
Eur. Phys. J. C \textbf{78}, no.3, 203 (2018)
%doi:10.1140/epjc/s10052-018-5662-y
[arXiv:1703.07364 [hep-ph]].
  %%CITATION = ARXIV:1703.07364;%%
  
%\cite{Roszkowski:2017nbc}
\bibitem{Roszkowski:2017nbc}
L.~Roszkowski, E.~M.~Sessolo and S.~Trojanowski,
%``WIMP dark matter candidates and searches?current status and future prospects,''
Rept. Prog. Phys. \textbf{81}, no.6, 066201 (2018)
%doi:10.1088/1361-6633/aab913
[arXiv:1707.06277 [hep-ph]].
 
%\cite{Kawasaki:2013ae}
\bibitem{Kawasaki:2013ae} 
  M.~Kawasaki and K.~Nakayama,
  %``Axions: Theory and Cosmological Role,''
  Ann.\ Rev.\ Nucl.\ Part.\ Sci.\  {\bf 63}, 69 (2013)
% doi:10.1146/annurev-nucl-102212-170536
  [arXiv:1301.1123 [hep-ph]].
  %%CITATION = doi:10.1146/annurev-nucl-102212-170536;%%

%\cite{Graham:2015ouw}
\bibitem{Graham:2015ouw} 
  P.~W.~Graham, I.~G.~Irastorza, S.~K.~Lamoreaux, A.~Lindner and K.~A.~van Bibber,
  %``Experimental Searches for the Axion and Axion-Like Particles,''
  Ann.\ Rev.\ Nucl.\ Part.\ Sci.\  {\bf 65}, 485 (2015)
%  doi:10.1146/annurev-nucl-102014-022120
  [arXiv:1602.00039 [hep-ex]].
  %%CITATION = doi:10.1146/annurev-nucl-102014-022120;%%

%\cite{Irastorza:2018dyq}
\bibitem{Irastorza:2018dyq}
I.~G.~Irastorza and J.~Redondo,
%``New experimental approaches in the search for axion-like particles,''
Prog. Part. Nucl. Phys. \textbf{102}, 89-159 (2018)
%doi:10.1016/j.ppnp.2018.05.003
[arXiv:1801.08127 [hep-ph]].


\bibitem{LHC}
  K. Pachal,  ``Dark Matter Searches at ATLAS and CMS'', given at the $8^{th}$ {\it {Edition of the Large Hadron Collider Physics Conference}}, 25-30 May, 2020.
 
 %\cite{Aprile:2018dbl}
\bibitem{Aprile:2018dbl}
E.~Aprile \textit{et al.} [XENON],
%``Dark Matter Search Results from a One Ton-Year Exposure of XENON1T,''
Phys. Rev. Lett. \textbf{121}, no.11, 111302 (2018)
%doi:10.1103/PhysRevLett.121.111302
[arXiv:1805.12562 [astro-ph.CO]]. 

%\cite{Fermi-LAT:2016uux}
\bibitem{Fermi-LAT:2016uux}
A.~Albert \textit{et al.} [Fermi-LAT and DES],
%``Searching for Dark Matter Annihilation in Recently Discovered Milky Way Satellites with Fermi-LAT,''
Astrophys. J. \textbf{834}, no.2, 110 (2017)
%doi:10.3847/1538-4357/834/2/110
[arXiv:1611.03184 [astro-ph.HE]].

%\cite{Amole:2019fdf}
\bibitem{Amole:2019fdf}
C.~Amole \textit{et al.} [PICO],
%``Dark Matter Search Results from the Complete Exposure of the PICO-60 C$_3$F$_8$ Bubble Chamber,''
Phys. Rev. D \textbf{100}, no.2, 022001 (2019)
%doi:10.1103/PhysRevD.100.022001
[arXiv:1902.04031 [astro-ph.CO]].


%\cite{Steigman:2015hda}
\bibitem{Steigman:2015hda} 
  G.~Steigman,
  %``CMB Constraints On The Thermal WIMP Mass And Annihilation Cross Section,''
  Phys.\ Rev.\ D {\bf 91}, no. 8, 083538 (2015)
%  doi:10.1103/PhysRevD.91.083538
  [arXiv:1502.01884 [astro-ph.CO]].
  %%CITATION = doi:10.1103/PhysRevD.91.083538;%% 
  
%\cite{Saikawa:2020swg}
\bibitem{Saikawa:2020swg}
K.~Saikawa and S.~Shirai,
%``Precise WIMP Dark Matter Abundance and Standard Model Thermodynamics,''
[arXiv:2005.03544 [hep-ph]].

%\cite{Alexander:2016aln}
\bibitem{Alexander:2016aln} 
  J.~Alexander {\it et al.},
  %``Dark Sectors 2016 Workshop: Community Report,''
  arXiv:1608.08632 [hep-ph].
  %%CITATION = ARXIV:1608.08632;%%

%\cite{Battaglieri:2017aum}
\bibitem{Battaglieri:2017aum} 
  M.~Battaglieri {\it et al.},
  %``US Cosmic Visions: New Ideas in Dark Matter 2017: Community Report,''
  arXiv:1707.04591 [hep-ph].
  %%CITATION = ARXIV:1707.04591;%%
  
 %\cite{Bertone:2018krk}
\bibitem{Bertone:2018krk}
G.~Bertone and T.~Tait, M.P.,
%``A new era in the search for dark matter,''
Nature \textbf{562}, no.7725, 51-56 (2018)
%doi:10.1038/s41586-018-0542-z
[arXiv:1810.01668 [astro-ph.CO]].
   

\bibitem{KM}
 %\cite{Holdom:1985ag}
%\bibitem{Holdom:1985ag} 
  B.~Holdom,
  %``Two U(1)'s and Epsilon Charge Shifts,''
  Phys.\ Lett.\  {\bf 166B}, 196 (1986) and
%  doi:10.1016/0370-2693(86)91377-8
  %%CITATION = doi:10.1016/0370-2693(86)91377-8;%%
%\cite{Holdom:1986eq}
%\bibitem{Holdom:1986eq} 
 %B.~Holdom,
  %``Searching for $\epsilon$ Charges and a New U(1),''
  Phys.\ Lett.\ B {\bf 178}, 65 (1986); 
%  doi:10.1016/0370-2693(86)90470-3
  %%CITATION = doi:10.1016/0370-2693(86)90470-3;%%
%\cite{Dienes:1996zr}
%\bibitem{Dienes:1996zr} 
  K.~R.~Dienes, C.~F.~Kolda and J.~March-Russell,
  %``Kinetic mixing and the supersymmetric gauge hierarchy,''
  Nucl.\ Phys.\ B {\bf 492}, 104 (1997)
%  doi:10.1016/S0550-3213(97)80028-4, 10.1016/S0550-3213(97)00173-9
  [hep-ph/9610479];
  %%CITATION = doi:10.1016/S0550-3213(97)80028-4, 10.1016/S0550-3213(97)00173-9;%%
 %\cite{DelAguila:1993px}
%\bibitem{DelAguila:1993px} 
  F.~Del Aguila,
  %``The Physics of z-prime bosons,''
  Acta Phys.\ Polon.\ B {\bf 25}, 1317 (1994)
  [hep-ph/9404323];
  %%CITATION = HEP-PH/9404323;%%
  %39 citations counted in INSPIRE as of 11 Jan 2018
%\cite{Babu:1996vt}
%\bibitem{Babu:1996vt} 
  K.~S.~Babu, C.~F.~Kolda and J.~March-Russell,
  %``Leptophobic U(1) $s$ and the R($b$) - R($c$) crisis,''
  Phys.\ Rev.\ D {\bf 54}, 4635 (1996)
%  doi:10.1103/PhysRevD.54.4635
  [hep-ph/9603212];
  %%CITATION = doi:10.1103/PhysRevD.54.4635;%%
  %200 citations counted in INSPIRE as of 11 Jan 2018
%\cite{Rizzo:1998ut}
%\bibitem{Rizzo:1998ut} 
  T.~G.~Rizzo,
  %``Gauge kinetic mixing and leptophobic $Z^\prime$ in E(6) and SO(10),''
  Phys.\ Rev.\ D {\bf 59}, 015020 (1998)
%  doi:10.1103/PhysRevD.59.015020
  [hep-ph/9806397].
  %%CITATION = doi:10.1103/PhysRevD.59.015020;%%
  %87 citations counted in INSPIRE as of 11 Jan 2018

\bibitem{vectorportal} 
 There has been a huge amount of work on this subject; see, for example, 
%\cite{Feldman:2006wd}
%\bibitem{Feldman:2006wd} 
  D.~Feldman, B.~Kors and P.~Nath,
  %``Extra-weakly Interacting Dark Matter,''
  Phys.\ Rev.\ D {\bf 75}, 023503 (2007)
  %doi:10.1103/PhysRevD.75.023503
  [hep-ph/0610133];
  %%CITATION = doi:10.1103/PhysRevD.75.023503;%%
  %94 citations counted in INSPIRE as of 26 Jan 2018
%\cite{Feldman:2007wj}
%\bibitem{Feldman:2007wj} 
  D.~Feldman, Z.~Liu and P.~Nath,
  %``The Stueckelberg Z-prime Extension with Kinetic Mixing and Milli-Charged Dark Matter From the Hidden Sector,''
  Phys.\ Rev.\ D {\bf 75}, 115001 (2007)
 % doi:10.1103/PhysRevD.75.115001
  [hep-ph/0702123 [HEP-PH]].;
  %%CITATION = doi:10.1103/PhysRevD.75.115001;%%
  %218 citations counted in INSPIRE as of 26 Jan 2018
 %\cite{Pospelov:2007mp}
%\bibitem{Pospelov:2007mp} 
  M.~Pospelov, A.~Ritz and M.~B.~Voloshin,
  %``Secluded WIMP Dark Matter,''
  Phys.\ Lett.\ B {\bf 662}, 53 (2008)
%  doi:10.1016/j.physletb.2008.02.052
  [arXiv:0711.4866 [hep-ph]];
  %%CITATION = doi:10.1016/j.physletb.2008.02.052;%%
  %500 citations counted in INSPIRE as of 18 Jan 2018
%\cite{Pospelov:2008zw}
%\bibitem{Pospelov:2008zw} 
  M.~Pospelov,
  %``Secluded U(1) below the weak scale,''
  Phys.\ Rev.\ D {\bf 80}, 095002 (2009)
%  doi:10.1103/PhysRevD.80.095002
  [arXiv:0811.1030 [hep-ph]]; 
  %%CITATION = doi:10.1103/PhysRevD.80.095002;%%
%\cite{Davoudiasl:2012qa}
%\bibitem{Davoudiasl:2012qa} 
  H.~Davoudiasl, H.~S.~Lee and W.~J.~Marciano,
  %``Muon Anomaly and Dark Parity Violation,''
  Phys.\ Rev.\ Lett.\  {\bf 109}, 031802 (2012)
%  doi:10.1103/PhysRevLett.109.031802
  [arXiv:1205.2709 [hep-ph]] and 
  %%CITATION = doi:10.1103/PhysRevLett.109.031802;%%
  %49 citations counted in INSPIRE as of 11 Jan 2018
%\cite{Davoudiasl:2012ag}
%\bibitem{Davoudiasl:2012ag} 
% H.~Davoudiasl, H.~S.~Lee and W.~J.~Marciano,
  %``'Dark' Z implications for Parity Violation, Rare Meson Decays, and Higgs Physics,''
  Phys.\ Rev.\ D {\bf 85}, 115019 (2012)
  doi:10.1103/PhysRevD.85.115019
  [arXiv:1203.2947 [hep-ph]];
  %%CITATION = doi:10.1103/PhysRevD.85.115019;%%
  %124 citations counted in INSPIRE as of 11 Jan 2018
  %\cite{Essig:2013lka}
%\bibitem{Essig:2013lka} 
  R.~Essig {\it et al.},
  %``Working Group Report: New Light Weakly Coupled Particles,''
  arXiv:1311.0029 [hep-ph];
  %%CITATION = ARXIV:1311.0029;%%
  %322 citations counted in INSPIRE as of 13 Jan 2018
%\cite{Izaguirre:2015yja}
%\bibitem{Izaguirre:2015yja} 
  E.~Izaguirre, G.~Krnjaic, P.~Schuster and N.~Toro,
  %``Analyzing the Discovery Potential for Light Dark Matter,''
  Phys.\ Rev.\ Lett.\  {\bf 115}, no. 25, 251301 (2015)
%  doi:10.1103/PhysRevLett.115.251301
  [arXiv:1505.00011 [hep-ph]];
  %%CITATION = doi:10.1103/PhysRevLett.115.251301;%%
  %35 citations counted in INSPIRE as of 13 Jan 2018
    %\cite{Khlopov:2013ava}
%\bibitem{Khlopov:2013ava} 
  M.~Khlopov,
  %``Fundamental Particle Structure in the Cosmological Dark Matter,''
  Int.\ J.\ Mod.\ Phys.\ A {\bf 28}, 1330042 (2013)
%  doi:10.1142/S0217751X13300421
  [arXiv:1311.2468 [astro-ph.CO]];
  %%CITATION = doi:10.1142/S0217751X13300421;%%
 For a general overview and introduction to this framework, see  
 %\cite{Curtin:2014cca}
%\bibitem{Curtin:2014cca} 
  D.~Curtin, R.~Essig, S.~Gori and J.~Shelton,
  %``Illuminating Dark Photons with High-Energy Colliders,''
  JHEP {\bf 1502}, 157 (2015)
%  doi:10.1007/JHEP02(2015)157
  [arXiv:1412.0018 [hep-ph]].
  %%CITATION = doi:10.1007/JHEP02(2015)157;%%
 
%\cite{Fabbrichesi:2020wbt}
\bibitem{Fabbrichesi:2020wbt}
M.~Fabbrichesi, E.~Gabrielli and G.~Lanfranchi,
%``The Dark Photon,''
[arXiv:2005.01515 [hep-ph]].

%\cite{Rizzo:2018vlb}
\bibitem{Rizzo:2018vlb}
T.~G.~Rizzo,
%``Kinetic Mixing and Portal Matter Phenomenology,''
Phys. Rev. D \textbf{99}, no.11, 115024 (2019)
%doi:10.1103/PhysRevD.99.115024
[arXiv:1810.07531 [hep-ph]].
%4 citations counted in INSPIRE as of 25 May 2020

%\cite{Rueter:2019wdf}
\bibitem{Rueter:2019wdf}
T.~D.~Rueter and T.~G.~Rizzo,
%``Towards A UV-Model of Kinetic Mixing and Portal Matter,''
Phys. Rev. D \textbf{101}, no.1, 015014 (2020)
%doi:10.1103/PhysRevD.101.015014
[arXiv:1909.09160 [hep-ph]].

%\cite{Kim:2019oyh}
\bibitem{Kim:2019oyh}
J.~H.~Kim, S.~D.~Lane, H.~S.~Lee, I.~M.~Lewis and M.~Sullivan,
%``Searching for Dark Photons with Maverick Top Partners,''
Phys. Rev. D \textbf{101}, no.3, 035041 (2020)
%doi:10.1103/PhysRevD.101.035041
[arXiv:1904.05893 [hep-ph]].

%\cite{Wojcik:2020wgm}
\bibitem{Wojcik:2020wgm}
G.~N.~Wojcik and T.~G.~Rizzo,
%``$SU(4)$ Flavorful Portals,''
[arXiv:2012.05406 [hep-ph]].

%\cite{Rueter:2020qhf}
\bibitem{Rueter:2020qhf}
T.~D.~Rueter and T.~G.~Rizzo,
%``Building Kinetic Mixing From Scalar Portal Matter,''
[arXiv:2011.03529 [hep-ph]].

%\cite{Rizzo:2020ybl}
\bibitem{Rizzo:2020ybl}
T.~G.~Rizzo and G.~N.~Wojcik,
%``Kinetic Mixing, Dark Photons and Extra Dimensions III: Brane Localized Dark Matter,''
[arXiv:2006.06858 [hep-ph]].

%\cite{Landim:2019epv}
\bibitem{Landim:2019epv}
R.~G.~Landim and T.~G.~Rizzo,
%``Thick Branes in Extra Dimensions and Suppressed Dark Couplings,''
JHEP \textbf{06}, 112 (2019)
%doi:10.1007/JHEP06(2019)112
[arXiv:1902.08339 [hep-ph]].

%\cite{Rizzo:2018joy}
\bibitem{Rizzo:2018joy}
T.~G.~Rizzo,
%``Kinetic mixing, dark photons and extra dimensions. Part II: fermionic dark matter,''
JHEP \textbf{10}, 069 (2018)
%doi:10.1007/JHEP10(2018)069
[arXiv:1805.08150 [hep-ph]].

%\cite{Rizzo:2018ntg}
\bibitem{Rizzo:2018ntg}
T.~G.~Rizzo,
%``Kinetic mixing, dark photons and an extra dimension. Part I,''
JHEP \textbf{07}, 118 (2018)
%doi:10.1007/JHEP07(2018)118
[arXiv:1801.08525 [hep-ph]].

%\cite{Sabti:2019mhn}
\bibitem{Sabti:2019mhn}
N.~Sabti, J.~Alvey, M.~Escudero, M.~Fairbairn and D.~Blas,
%``Refined Bounds on MeV-scale Thermal Dark Sectors from BBN and the CMB,''
JCAP \textbf{01}, 004 (2020)
%doi:10.1088/1475-7516/2020/01/004
[arXiv:1910.01649 [hep-ph]].
 
  
%\cite{Slatyer:2015jla}
\bibitem{Slatyer:2015jla}
T.~R.~Slatyer,
%``Indirect dark matter signatures in the cosmic dark ages. I. Generalizing the bound on s-wave dark matter annihilation from Planck results,''
Phys. Rev. D \textbf{93}, no.2, 023527 (2016)
%doi:10.1103/PhysRevD.93.023527
[arXiv:1506.03811 [hep-ph]].

%\cite{Liu:2016cnk}
\bibitem{Liu:2016cnk} 
  H.~Liu, T.~R.~Slatyer and J.~Zavala,
  %``Contributions to cosmic reionization from dark matter annihilation and decay,''
  Phys.\ Rev.\ D {\bf 94}, no. 6, 063507 (2016)
%  doi:10.1103/PhysRevD.94.063507
  [arXiv:1604.02457 [astro-ph.CO]].
  %%CITATION = doi:10.1103/PhysRevD.94.063507;%%

%\cite{Leane:2018kjk}
\bibitem{Leane:2018kjk}
R.~K.~Leane, T.~R.~Slatyer, J.~F.~Beacom and K.~C.~Ng,
%``GeV-scale thermal WIMPs: Not even slightly ruled out,''
Phys. \ Rev. \ D \textbf{98}, no.2, 023016 (2018)
%doi:10.1103/PhysRevD.98.023016
[arXiv:1805.10305 [hep-ph]].

%\cite{Bringmann:2006mu}
\bibitem{Bringmann:2006mu}
T.~Bringmann and S.~Hofmann,
%``Thermal decoupling of WIMPs from first principles,''
JCAP \textbf{04}, 016 (2007)
[erratum: JCAP \textbf{03}, E02 (2016)]
%doi:10.1088/1475-7516/2007/04/016
[arXiv:hep-ph/0612238 [hep-ph]].

%\cite{Cang:2020exa}
\bibitem{Cang:2020exa}
J.~Cang, Y.~Gao and Y.~Z.~Ma,
%``Probing Dark Matter with Future CMB Measurements,''
[arXiv:2002.03380 [astro-ph.CO]].

%\cite{Green:2018pmd}
\bibitem{Green:2018pmd}
D.~Green, P.~D.~Meerburg and J.~Meyers,
%``Aspects of Dark Matter Annihilation in Cosmology,''
JCAP \textbf{04}, 025 (2019)
%doi:10.1088/1475-7516/2019/04/025
[arXiv:1804.01055 [astro-ph.CO]].

%\cite{Ade:2018sbj}
\bibitem{Ade:2018sbj}
P.~Ade \textit{et al.} [Simons Observatory],
%``The Simons Observatory: Science goals and forecasts,''
JCAP \textbf{02}, 056 (2019)
%doi:10.1088/1475-7516/2019/02/056
[arXiv:1808.07445 [astro-ph.CO]].

%\cite{Abazajian:2016yjj}
\bibitem{Abazajian:2016yjj}
K.~N.~Abazajian \textit{et al.} [CMB-S4],
%``CMB-S4 Science Book, First Edition,''
[arXiv:1610.02743 [astro-ph.CO]].

%\cite{Boudaud:2016mos}
\bibitem{Boudaud:2016mos}
M.~Boudaud, J.~Lavalle and P.~Salati,
%``Novel cosmic-ray electron and positron constraints on MeV dark matter particles,''
Phys. Rev. Lett. \textbf{119}, no.2, 021103 (2017)
%doi:10.1103/PhysRevLett.119.021103
[arXiv:1612.07698 [astro-ph.HE]].

%\cite{Boudaud:2018oya}
\bibitem{Boudaud:2018oya}
M.~Boudaud, T.~Lacroix, M.~Stref and J.~Lavalle,
%``Robust cosmic-ray constraints on $p$-wave annihilating MeV dark matter,''
Phys. \ Rev. \ D \textbf{99}, no.6, 061302 (2019)
%doi:10.1103/PhysRevD.99.061302
[arXiv:1810.01680 [astro-ph.HE]].


%\cite{Griest:1990kh}
\bibitem{Griest:1990kh}
K.~Griest and D.~Seckel,
%``Three exceptions in the calculation of relic abundances,''
Phys. Rev. D \textbf{43}, 3191-3203 (1991).
%doi:10.1103/PhysRevD.43.3191

%\cite{DAgnolo:2015ujb}
\bibitem{DAgnolo:2015ujb}
R.~T.~D'Agnolo and J.~T.~Ruderman,
%``Light Dark Matter from Forbidden Channels,''
Phys. Rev. Lett. \textbf{115}, no.6, 061301 (2015)
%doi:10.1103/PhysRevLett.115.061301
[arXiv:1505.07107 [hep-ph]].

%\cite{Cline:2017tka}
\bibitem{Cline:2017tka}
J.~M.~Cline, H.~Liu, T.~Slatyer and W.~Xue,
%``Enabling Forbidden Dark Matter,''
Phys. Rev. D \textbf{96}, no.8, 083521 (2017)
%doi:10.1103/PhysRevD.96.083521
[arXiv:1702.07716 [hep-ph]].

%\cite{Fitzpatrick:2020vba}
\bibitem{Fitzpatrick:2020vba}
P.~J.~Fitzpatrick, H.~Liu, T.~R.~Slatyer and Y.~D.~Tsai,
%``New Pathways to the Relic Abundance of Vector-Portal Dark Matter,''
[arXiv:2011.01240 [hep-ph]].

%\cite{1837855}
\bibitem{1837855}
R.~T.~D'Agnolo, D.~Liu, J.~T.~Ruderman and P.~J.~Wang,
%``Forbidden Dark Matter Annihilations into Standard Model Particles,''
[arXiv:2012.11766 [hep-ph]].

%\cite{Berlin:2014tja}
\bibitem{Berlin:2014tja}
A.~Berlin, D.~Hooper and S.~D.~McDermott,
%``Simplified Dark Matter Models for the Galactic Center Gamma-Ray Excess,''
Phys. Rev. D \textbf{89}, no.11, 115022 (2014)
%doi:10.1103/PhysRevD.89.115022
[arXiv:1404.0022 [hep-ph]].


%\cite{Plehn:2017fdg}
\bibitem{Plehn:2017fdg}
M.~Bauer and T.~Plehn,
%``Yet Another Introduction to Dark Matter: The Particle Physics Approach,''
Lect. Notes Phys. \textbf{959}, pp. (2019)
%doi:10.1007/978-3-030-16234-4
[arXiv:1705.01987 [hep-ph]].

%\cite{Feng:2017drg}
\bibitem{Feng:2017drg} 
J.~L.~Feng and J.~Smolinsky,
  %``Impact of a resonance on thermal targets for invisible dark photon searches,''
  Phys.\ Rev.\ D {\bf 96}, no. 9, 095022 (2017)
%  doi:10.1103/PhysRevD.96.095022
  [arXiv:1707.03835 [hep-ph]] .
  %%CITATION = doi:10.1103/PhysRevD.96.095022;%%

%\cite{Li:2015tka}
\bibitem{Li:2015tka} 
  B.~Li and Y.~F.~Zhou,
  %``Direct detection of dark matter with resonant annihilation,''
  Commun.\ Theor.\ Phys.\  {\bf 64}, no. 1, 119 (2015)
%  doi:10.1088/0253-6102/64/1/119
  [arXiv:1503.08281 [hep-ph]]. 
  %%CITATION = doi:10.1088/0253-6102/64/1/119;%%
  
%\cite{Bernreuther:2020koj}
\bibitem{Bernreuther:2020koj}
E.~Bernreuther, S.~Heeba and F.~Kahlhoefer,
%``Resonant Sub-GeV Dirac Dark Matter,''
[arXiv:2010.14522 [hep-ph]].
  
 %\cite{Duch:2018ucs}
\bibitem{Duch:2018ucs}
M.~Duch, B.~Grzadkowski and A.~Pilaftsis,
%``Gauge-Independent Approach to Resonant Dark Matter Annihilation,''
JHEP \textbf{02}, 141 (2019)
%doi:10.1007/JHEP02(2019)141
[arXiv:1812.11944 [hep-ph]].
 
%\cite{Das:2020rxn}
\bibitem{Das:2020rxn}
A.~Das, K.~Enomoto and S.~Kanemura,
%``Alternative dark matter phenomenology in a general $U(1)_X$ extension of the Standard Model,''
[arXiv:2011.04537 [hep-ph]].

%\cite{Rizzo:2020jsm}
\bibitem{Rizzo:2020jsm}
T.~G.~Rizzo,
%``Dark Initial State Radiation and the Kinetic Mixing Portal,''
[arXiv:2006.08502 [hep-ph]].

%\cite{Aguilar-Arevalo:2019wdi}
\bibitem{Aguilar-Arevalo:2019wdi}
A.~Aguilar-Arevalo \textit{et al.} [DAMIC],
%``Constraints on Light Dark Matter Particles Interacting with Electrons from DAMIC at SNOLAB,''
Phys. Rev. Lett. \textbf{123}, no.18, 181802 (2019)
%doi:10.1103/PhysRevLett.123.181802
[arXiv:1907.12628 [astro-ph.CO]].

%\cite{Aprile:2019xxb}
\bibitem{Aprile:2019xxb}
E.~Aprile \textit{et al.} [XENON],
%``Light Dark Matter Search with Ionization Signals in XENON1T,''
Phys. Rev. Lett. \textbf{123}, no.25, 251801 (2019)
%doi:10.1103/PhysRevLett.123.251801
[arXiv:1907.11485 [hep-ex]].

%\cite{Aprile:2020tmw}
\bibitem{Aprile:2020tmw}
E.~Aprile \textit{et al.} [XENON],
%``Excess electronic recoil events in XENON1T,''
Phys. Rev. D \textbf{102}, no.7, 072004 (2020)
%doi:10.1103/PhysRevD.102.072004
[arXiv:2006.09721 [hep-ex]].

%\cite{Amaral:2020ryn}
\bibitem{Amaral:2020ryn}
D.~W.~Amaral \textit{et al.} [SuperCDMS],
%``Constraints on low-mass, relic dark matter candidates from a surface-operated SuperCDMS single-charge sensitive detector,''
Phys. Rev. D \textbf{102}, no.9, 091101 (2020)
%doi:10.1103/PhysRevD.102.091101
[arXiv:2005.14067 [hep-ex]].

%\cite{Arnaud:2020svb}
\bibitem{Arnaud:2020svb}
Q.~Arnaud \textit{et al.} [EDELWEISS],
%``First germanium-based constraints on sub-MeV Dark Matter with the EDELWEISS experiment,''
Phys. Rev. Lett. \textbf{125}, no.14, 141301 (2020)
%doi:10.1103/PhysRevLett.125.141301
[arXiv:2003.01046 [astro-ph.GA]].

%\cite{Barak:2020fql}
\bibitem{Barak:2020fql}
L.~Barak \textit{et al.} [SENSEI],
%``SENSEI: Direct-Detection Results on sub-GeV Dark Matter from a New Skipper-CCD,''
Phys. Rev. Lett. \textbf{125}, no.17, 171802 (2020)
%doi:10.1103/PhysRevLett.125.171802
[arXiv:2004.11378 [astro-ph.CO]].


\end{thebibliography}
\end{document}